\documentclass{pasj00}

\draft
\SetRunningHead{Y. Taniguchi}{Lyman$alpha$ Emitters at $z$=6.6}
\Received{2004/7/27}
\Accepted{2004/11/30}
\begin{document}
\title{The SUBARU Deep Field Project: LYMAN$\alpha$ Emitters at 
       a Redshift of 6.6\footnotemark[1] \footnotemark[2]
      }

\author{
          Yoshiaki \textsc{Taniguchi}     \altaffilmark{1},
          Msaru \textsc{Ajiki}            \altaffilmark{1},
          Tohru \textsc{Nagao}            \altaffilmark{1, 2},
          Yasuhiro \textsc{Shioya}        \altaffilmark{1},\\
          Takashi \textsc{Murayama}       \altaffilmark{1},
          Nobunari \textsc{Kashikawa}     \altaffilmark{3},
          Keiichi \textsc{Kodaira}        \altaffilmark{4},
          Norio \textsc{Kaifu}            \altaffilmark{3},\\
          Hiroyasu \textsc{Ando}          \altaffilmark{3},
          Hiroshi \textsc{Karoji}         \altaffilmark{5},
          Masayuki \textsc{Akiyama}       \altaffilmark{5},
          Kentaro \textsc{Aoki}           \altaffilmark{5},\\
          Mamoru \textsc{Doi}             \altaffilmark{6},
          Shinobu S. \textsc{Fujita}      \altaffilmark{1},
          Hisanori \textsc{Furusawa}      \altaffilmark{5},
          Tomoki \textsc{Hayashino}       \altaffilmark{7},\\
          Fumihide \textsc{Iwamuro}       \altaffilmark{8},
          Masanori \textsc{Iye}           \altaffilmark{3},
          Naoto \textsc{Kobayashi}        \altaffilmark{6},
          Tadayuki \textsc{Kodama}        \altaffilmark{3},\\
          Yutaka \textsc{Komiyama}        \altaffilmark{5},
          Yuichi \textsc{Matsuda}         \altaffilmark{3, 7},
          Satoshi \textsc{Miyazaki}       \altaffilmark{5},
          Yoshihiko \textsc{Mizumoto}     \altaffilmark{3},\\
          Tomoki \textsc{Morokuma}        \altaffilmark{6},
          Kentaro \textsc{Motohara}       \altaffilmark{6}, 
          Kyoji \textsc{Nariai}           \altaffilmark{9},
          Koji \textsc{Ohta}              \altaffilmark{8},\\
          Youichi \textsc{Ohyama}         \altaffilmark{5},
          Sadanori \textsc{Okamura}       \altaffilmark{10, 11},
          Masami \textsc{Ouchi}           \altaffilmark{3},
          Toshiyuki \textsc{Sasaki}       \altaffilmark{5},\\
          Yasunori \textsc{Sato}          \altaffilmark{3},
          Kazuhiro \textsc{Sekiguchi}     \altaffilmark{5},
          Kazuhiro \textsc{Shimasaku}     \altaffilmark{10},
          Hajime \textsc{Tamura}          \altaffilmark{7},\\
          Masayuki \textsc{Umemura}       \altaffilmark{12},
          Toru \textsc{Yamada}            \altaffilmark{3}, 
          Naoki \textsc{Yasuda}           \altaffilmark{13}, \& 
          Michitoshi  \textsc{Yoshida}    \altaffilmark{10},
}

\footnotetext[1]{Based on data collected at the Subaru Telescope, which is
                 operated by the National Astronomical Observatory of Japan.}
\footnotetext[2]{This work has been done with a collaboration between
       the Subaru Deep Field Project led by the association of
       builders of the Subaru telescope and the common-use,
       Intensive Program (S02A-IP-2) led by Y. Taniguchi.}

\altaffiltext{1}{Astronomical Institute, Graduate School of Science,
         Tohoku University, Aramaki, Aoba, Sendai 980-8578}
\altaffiltext{2}{INAF -- Osservatorio Astrofisico di Arcetri,
         Largo Enrico Fermi 5, 50125 Firenze, Italy}
\altaffiltext{3}{National Astronomical Observatory of Japan,
          2-21-1 Osawa, Mitaka, Tokyo 181-8588}
\altaffiltext{4}{The Graduate University for Advanced Studies,
        Shonan Village, Hayama, Kanagawa 240-0193}
\altaffiltext{5}{Subaru Telescope, National Astronomical Observatory of Japan,\\
          650 N. A'ohoku Place, Hilo, HI 96720, USA}
\altaffiltext{6}{Institute of Astronomy, Graduate School of Science,\\
          The University of Tokyo, 2-21-1 Osawa, Mitaka, Tokyo 181-0015}
\altaffiltext{7}{Research Center for Neutrino Science, Graduate School of Science,\\
        Tohoku University, Aramaki, Aoba, Sendai 980-8578}
\altaffiltext{8}{Department of Astronomy,  Graduate School of Science,\\
        Kyoto University, Kitashirakawa, Sakyo, Kyoto 606-8502}
\altaffiltext{9}{Department of Physics, Meisei University, 2-1-1 Hodokubo,
        Hino, Tokyo 191-8506}
\altaffiltext{10}{Department of Astronomy, Graduate School of Science,
        The University of Tokyo, Tokyo 113-0033}
\altaffiltext{11}{Research Center for the Early Universe, Graduate School\\
        of Science, The University of Tokyo, Tokyo 113-0033}
\altaffiltext{12}{Center for Computational Physics, University of Tsukuba,
        1-1-1 Tennodai, Tsukuba 305-8571}
\altaffiltext{13}{Institute for Cosmic Ray Research, The University of Tokyo,
        Kashiwa, 277-8582}


\maketitle

\begin{abstract}
We present new results of a deep optical imaging survey using a narrow band
filter ($NB921$) centered at $\lambda =$ 9196 \AA ~ together with $B$, $V$,
$R$, $i^\prime$, and $z^\prime$  broadband filters in the sky area of
the Subaru Deep Field, which has been promoted as one of legacy programs
of the 8.2 m Subaru Telescope.
We obtained a photometric sample of 58 Ly$\alpha$ emitter candidates at $z \approx$ 6.5 --
6.6 among $\sim 180$ strong  $NB921$-excess ($z^\prime - NB921 > 1.0$) objects 
together with a color criterion of $i^\prime - z^\prime > 1.3$.
We then obtained optical spectra of 20 objects 
in our $NB921$-excess sample, and identified at least nine  
Ly$\alpha$ emitters at $z \sim 6.5$ -- 6.6, including the two emitters
reported by Kodaira et al. (2003, PASJ, 55, L17). 
Since our Ly$\alpha$-emitter 
candidates are free from strong amplification of gravitational
lensing, we are able to discuss their observational properties from a statistical
point of view. 
Based on these new results, we obtained a lower limit of the 
star-formation rate density of
$\rho_{\rm SFR} \simeq 5.7 \times 10^{-4}$ $h_{0.7}$ $M_\odot$ yr$^{-1}$  Mpc$^{-3}$
at $z \approx 6.6$, being consistent with our previous estimate.
We discuss the nature of star-formation activity in galaxies beyond $z=6$.
\end{abstract}

\section{INTRODUCTION}

A simple method to understand the formation process of galaxies is
to find a sample of very young galaxies at high redshift, and
then to investigate their observational properties in detail.
Theoretical models based on the hierarchical clustering scenario
suggest that first-generation (i.e., Population III) objects 
could be born around $z \sim 30$ ($\sim$ 0.5 Gyr after the big bang),
and then galactic systems with masses higher than $10^{10} M_\odot$
could be assembled after $z \sim 5$ -- 10 ($\sim$ 1 Gyr after the
big bang) although their co-moving number density could be significantly smaller
than that in the present day  (e.g., Ostriker, Gnedin 1996). 
It is, thus, important 
to search for star-forming galaxies beyond redshift 5 to probe the
cosmic star-formation history from very high redshift to the present day.

Surveys for such high-$z$ galaxies have been made mostly by the optical
color-selection technique [e.g., Steidel et al. (1996), Madau et al. (1996)
see for the recent discovery of a galaxy at $z=10$, Pell\'o et al. 
(2004a); note that the detection of Ly$\alpha$ emission line has been still
in debate; see Weatherley et al. (2004); Pell\'o et al. (2004b)].
Indeed, recent wide-field, deep imaging surveys using the Advanced Camera
for Surveys on board the Hubble Space Telescope surely found a sample of
galaxies at $z \sim 6$ (Giavalisco et al. 2004b; Dickinson et al. 2004;
Bouwens et al. 2004; Stanway et al. 2004a, 2004b).
In addition, recent advances in deep optical imaging
capability with 8--10 m class telescopes
have enabled new searches for star-forming galaxies beyond redshift 5
(see for review, Taniguchi et al. 2003b; Spinrad 2003).
In particular, imaging surveys using narrow-passband filters have
proved to be an efficient way to find such galaxies (e.g., Hu et al. 2002,
2004; Kodaira et al. 2003; Maier et al. 2003; Rhoads et al. 2003, 2004).
These surveys can probe objects with a high star-formation rate, 
even if they are too faint to be detected in color-selection procedures.
Therefore, these two methods cooperate with each other to investigate
young, star-forming galaxies at high redshift. It is also worthwhile
noting that recent imaging-spectroscopic 
surveys with a grism are also capable of finding high-$z$ LAEs. A recent
such trial succeeded to find a LAE at $z = 6.518$ (Kurk et al. 2004).
A serendipitous discovery of a LAE at $z = 6.545$ was  also reported 
(Stern et al. 2004). For a review of discovery methods, see Taniguchi 
et al. (2003b).

The Subaru Telescope team
officially started a large-scale deep-survey program in 2002 April;
the Subaru Deep Field (SDF) project.
Several pilot papers related to the SDF project have already been published
(Maihara et al. 2001; Totani et al. 2001a, 2001b, 2001c; Ouchi et al. 2001, 2003,
2004a, 2004b; Shimasaku et al. 2003; Kashikawa et al. 2003).
The main aim of this project is to investigate the formation and evolution of
galaxies from very high redshift ($z \sim7$) to the present day ($z \sim 0$)
based on a set of very deep optical and near infrared (NIR) imaging data and
follow-up spectroscopy in both the optical and NIR. In addition, other
multi-wavelength observations will be added in the future.
The overall design of the SDF project will be given elsewhere
(Kashikawa et al. 2004a).

Since the SDF data set is very deep in both the broad and narrow filter bands
(see table 1), we will be able to contribute to the progress in searches for such
high-$z$ galaxies based on the above two methods. Using our SDF data taken 
in 2002, we have already found two Ly$\alpha$ emitters (LAEs)
at $z=6.578$ and $z=6.541$ (Kodaira et al. 2003). The former object is
the most distant Ly$\alpha$ emitter found in optical surveys, SDF J132418.3+271455;
note that the second most distant known is HCM 6A at $z=6.56$ (Hu et al. 2002).
In this paper, we present a summary of our search for LAEs at $z \approx 6.6$
made during these two years.
We adopt a flat universe with $\Omega_{\rm matter} = 0.3$,
$\Omega_{\Lambda} = 0.7$, and $h_{0.7} = H_0/(70 ~ {\rm km ~ s}^{-1}$
Mpc$^{-1}$) throughout this paper.
We use the AB system for optical magnitudes.

\section{OBSERVATIONS}

\subsection{Optical Deep Imaging}

We have carried out a very deep optical imaging survey in the SDF
centered at $\alpha$(J2000) = $13^{\rm h} ~ 24^{\rm m} ~ 38^{\rm s}.9$ and
$\delta$(J2000) = $+27^\circ ~ 29' ~ 25''.9$ (Kashikawa et al. 2004a).
Optical imaging was made in the $B$, $V$, $R$, $i'$, $z'$, $NB816$,
and  $NB921$ bands
on a central $34'\times 27'$ area of the SDF with Suprime-Cam,
which consisted of $5\times 2$ CCDs of 2k $\times$ 4k pixels,
with a pixel scale of $0.''202$ pixel$^{-1}$ (Miyazaki et al. 2002)
on the 8.2 m Subaru Telescope (Kaifu et al. 2000; Iye et al. 2004).
The data were collected in several observing runs during a period
between 2001 and 2003. 
A summary of the imaging observations is given in table 1.

Two narrow band filters were used to find LAEs at
$z \approx 5.7$, and $z \approx 6.6$. 
Other narrow band filters,
$NB704$ and $NB711$, were also used in the SDF 
to map the SDF. Early results with the use of
$NB711$ were reported by Ouchi et al. (2003) and Shimasaku et al.
(2003), and those with the use of $NB704$ were by Shimasaku et al. (2004).
In this paper,
we present results obtained with the use of $NB921$ filter centered on
$\lambda_{\rm c}$ = 9196 \AA ~ with a passband of $\Delta\lambda$(FWHM) =
132 \AA, together with broad-band data.
The central wavelength corresponds to a redshift of 6.56 for
Ly$\alpha$ emission.
Details of the filter system used in the SDF project are given in
Kashikawa et al. (2004a).
The central wavelength corresponds to a redshift of 6.56 for
Ly$\alpha$ emission.
Results obtained by using our NB816 data will be reported elsewhere.

The individual CCD data were reduced and
combined using both our own data-reduction software
(Yagi et al. 2002) and IRAF.
The combined images for individual bands were aligned and
smoothed with Gaussian kernels to match their seeing sizes.
The PSF FWHM of the final images is $0.''98$.
The exposure times and limiting magnitudes are listed in table 1.
Photometric calibrations were made
using the usual spectrophotometric standard stars.
Source detection and photometry were performed using
SExtractor  (Bertin, Arnouts 1996) version 2.1.6.
The $NB921$-band image was chosen to detect candidate objects,
and we limit the object catalog to $NB921 \leq 26.54$,
the $3\sigma$ limiting magnitude.
For each object detected in the $NB921$ image, the
$i^\prime$, $z^\prime$, and $NB921$ magnitudes were measured on a common
aperture of $2.''0$ diameter.
In total, $\sim  82000$ objects were detected down to $NB921=26.54$.

The bandpass of the NB921 filter is completely included in that of the
$z^\prime$ filter. Therefore, when the Lyman break
($\lambda_{\rm rest}$ = 912 \AA), or nearly zero continuum flux at 
$\lambda_{\rm rest}$(Ly$\alpha$) = 1216 \AA ~ is redshifted to the $z^\prime$
window, Lyman-break galaxies could show some excess in the $z^\prime - NB921$
color, even if they have little Ly$\alpha$ emission.
In order to examine this possibility, in figure 1, 
we show the $z^\prime - NB921$ color as a function of the redshift for 
a galaxy without Ly$\alpha$ emission;
the model galaxy is generated by using GALAXEV (Bruzual, Charlot 2003)
with parameters of $Z=0.02$, $\tau =$ 1 Gyr, and the age = 1 Gyr.
Even for such a Lyman break galaxy without Ly$\alpha$ emission, 
the  $z^\prime - NB921$ color becomes as high as 0.9 mag at $z \sim 6.4$.
Therefore, we required that the $z^\prime - NB921$ color exceeds 1.0 mag
for strong Ly$\alpha$ emitters at $z\simeq 6.6$; i.e.,


\begin{equation}
z^\prime - NB921 > 1.0.
\end{equation}
Together with this criterion, we also used the following two more
selection criteria to select $NB921$-excess objects:

\begin{equation}
z^\prime - NB921 > 3 \sigma 
\end{equation}
and 

\begin{equation}
 NB921 < 26.0 (5\sigma). 
\end{equation}
Although our 3$\sigma$ detection limit for $NB921$ was 26.54,
we adopted the above 5$\sigma$ limit to secure the selection of
$NB921$-excess objects.
Using these criteria, we obtain a sample of 185 NB921-excess objects.
In figure 2, we show a diagram between $z' - NB921$ and $NB921$,
where these criteria are also shown.

In order to reduce contamination from
foreground objects that are free from absorption
by the intergalactic medium, we adopted another color criterion,

\begin{equation}
i^\prime - z^\prime > 1.3,
\end{equation}
together with that LAE candidates are not detected in $B$, $V$, and $R$
band images (less than $3 \sigma$ in each band).
In order to show this color criterion,
we plot a diagram between  $z' - NB921$ and $i' - z'$ in figure 3.
We also show the color evolution of a model galaxy as a function of the redshift;
colors of a starburst galaxy were calculated by using the population synthesis
model GALAXEV (Bruzual, Charlot 2003) where we adopted the $\tau=1$ Gyr model with
an age of $t=1$ Gyr and adding some emission lines, e.g., Ly$\alpha$, [O {\sc ii}],
and [O {\sc iii}]. The emission-line luminosities were calculated in the  
following way:
(i) the H$\beta$ luminosity was calculated from the ionizing photon production
rate, $N_{\rm Lyc}$ photons s$^{-1}$,  using the relation

\begin{equation}
L({\rm H \beta}) ({\rm erg \; s^{-1}}) = 4.76 \times 10^{-13} N_{\rm
Lyc} ({\rm s^{-1}}),
\end{equation}
and (ii) the other emission-line luminosities  are calculated
using the relative luminosity to H$\beta$ luminosity
tabulated in PEGASE (Fioc, Rocca-Volmerange 1997).



In order not to miss possible faint LAEs,
we included all objects with  $z'-NB921>1.0$ even fainter
than $i' = 27.87$ ($\simeq 2\sigma$ limiting mag) in our LAE sample.
Part of them nominally had $i'-z'>1.3$,
although we did not use the $i'-z'$ color criterion for
objects fainter than $z' \simeq 26.57$ ($= 27.87 - 1.3$).
By using all of these criteria, we obtained
 a photometric sample of 58 LAE candidates. 
Their basic properties are given in table 2.
In this table, we also give a pure $z^\prime$-band magnitude, $z_{9500}$, 
for each object.
This magnitude was corrected for the contribution of the Ly$\alpha$ flux, estimated
by using the $NB921$ magnitude. The corresponding pure $z^\prime$-band flux was estimated 
as

\begin{equation}
f_{9500} = [\Delta\lambda_{z^\prime} f(z^\prime) - 0.7 
\Delta\lambda_{NB921} f(NB921)]/\Delta\lambda_{z^\prime}^{\rm eff},
\end{equation}
where $f(z^\prime)$ and $f(NB921)$ are the observed $z^\prime$ and $NB921$ fluxes,
respectively,  $\Delta\lambda_{z^\prime}$ is the  bandpass of the $z^\prime$
filter in units of \AA~
(= 960 \AA), $\Delta\lambda_{NB921}$ is that of the NB921 filter
(= 132 \AA), $\Delta\lambda_{z^\prime}^{\rm eff}$ is the effective 
bandpass of the $z^\prime$ filter for $\lambda_{\rm obs} \geq$ 9260 \AA ~
(=270 \AA), and the numerical factor of 0.7 is
the relative transmittance of the NB921 filter with respect to the $z^\prime$ filter.

The Ly$\alpha$ flux corrected for the contribution of UV continuum 
emission at wavelengths longer than 1216 \AA ~ for each object was
estimated by using 

\begin{equation}
f_{\rm Ly\alpha, image} = f(NB921) - f_\lambda(z_{9500}) \times
\Delta\lambda_{NB921}/2.
\end{equation}

Here, we note that this sample was slightly different from that discussed in
Kodaira et al. (2003) and Taniguchi (2003), because the
photometric data used in their papers were based on data 
obtained before 2002 May and their LAE selection criteria 
were also different from those used in this paper.
The previous LAE selection criteria adopted in Kodaira et al.
(2003) were: (1) $z^\prime - NB921 > 1.0$ and
$i^\prime - z^\prime > 1.3$ for objects with $i^\prime > 28.0$,
and (2) $z^\prime - NB921 > 1.0$ for objects with $i^\prime 
\leq 28.0$. Note that Taniguchi (2003) adopted $z^\prime - NB921 > 0.9$
because the photometric sample at that occasion was tentatively selected for
our follow-up optical spectroscopy. 
Note that the photometric catalog used in this paper is the final one
described in Kashikawa et al. (2004a) and thus the selection procedure of
$NB921$-excess objects is also our final one. 

The observed equivalent width of Ly$\alpha$ emission for each galaxy 
in our photometric sample is given in table 2.
Our NB921-excess criterion given in (1) nominally corresponds
to the cutoff for the Ly$\alpha$-emission equivalent width,
$EW_{\rm obs}({\rm Ly}\alpha)>200$ \AA. However, 
since the actual equivalent width of
Ly$\alpha$ emission should be obtained from
$EW_{\rm obs}({\rm Ly}\alpha) = f_{\rm Ly\alpha, image}/f_{9500}$,
the NB921-excess criterion given in (1) gives
the equivalent width cutoff of $EW_{\rm obs}({\rm Ly}\alpha)>50$ \AA,
corresponding to the rest-frame equivalent width, $EW_0({\rm Ly}\alpha) > 7$ \AA.
However, most LAE candidates in our sample tend to be faint
in the $z^\prime$ band. Further, in order to secure the selection 
of LAE candidates, we also adopted the criterion (2): $z^\prime - NB921 >
3 \sigma$.  Therefore, LAEs with much stronger Ly$\alpha$
emission could be detected in our analysis; our nine LAEs have
$EW_{\rm obs}({\rm Ly}\alpha)> 130$ \AA ~ (see table 2).

The effective area used to search for $NB921$-excess objects was 875.4 arcmin$^2$.
The FWHM half-power points of the filter corresponded to a co-moving depth along
the line of sight of 40.9 $h_{0.7}^{-1}$ Mpc ($z_{\rm min} \approx
6.508$ and $z_{\rm max} \approx 6.617$; note that the transmission
curve of our $NB921$ filter has a Gaussian-like shape).  Therefore, a
total volume of 217200 $h_{0.7}^{-3}$ Mpc$^{3}$ was probed in our $NB921$ image.

\subsection{Optical Spectroscopy}

In order to investigate the nature of LAE candidates found in our
optical-imaging survey, we obtained optical spectra of 20
objects in our photometric sample of LAEs;
note that the nine galaxies for which optical spectroscopy was
made were randomly selected from a photometric sample.
We used the Subaru Faint Object Camera And Spectrograph (FOCAS; Kashikawa 
et al. 2002) on 2002 June and 2003 May and June.
A journal of our spectroscopy is summarized in table 3.
Details of this spectroscopy will be given elsewhere.

Our spectroscopy was made with the multi-object slit (MOS) mode using 
the following grating sets: (i) a 300 lines mm$^{-1}$ grating 
with an O58 order cut filter, and (ii) an Echelle with a $z^\prime$ filter.
An 0.$''$8-wide slit was used in both settings
The wavelength coverage was $\sim$ 6000 \AA ~ to 10000 \AA ~
for case (i), while it was $\sim$ 8000 \AA ~ to 10000 \AA ~
for case (ii). The spectroscopic resolution was
9.0 \AA ~ at 9200 \AA, $R \simeq 1020$, for case  (i)
while 6.3 \AA ~ at 9000 \AA,  $R \simeq 1430$, for case (ii).
The spatial sampling rate was 0.$^{\prime\prime}$3 pixel$^{-1}$
after 3-pixel, on-chip binning in both cases.
The flux calibration was made with spectra of spectroscopic standard stars,
Hz 44 and Feige 34.

In figure 4, we show the distributions of the $NB921$ magnitudes for
the photometric sample (58 objects) and the spectroscopic sample (20 objects);
note that those of the spectroscopically-confirmed LAE sample (9 objects),
given in subsection 3.1, are also shown.
In figure 5, we show the spatial distributions of the above three samples.
Details on this issue will be discussed by Kashikawa et al. (2004b,
in preparation).



\section{RESULTS}

\subsection{Spectroscopic Identification}

Our photometric sample of LAE candidates consisted of 58 objects. 
We obtained optical spectra of 20 objects among them.
We detected single emission lines in 14 objects, but did not detected
any emission-line feature in the  remaining six objects,
although one shows a marginal continuum feature.

The fourteen emission-line objects with a single emission line at 
$\lambda \simeq$ 9200 \AA ~  may be either an LAE at $z \approx 6.6$
or an [O {\sc ii}] emitter at $z \approx 1.47$ (e.g., Stern et al. 2000).
LAEs at high redshift show a sharp cutoff at wavelengths shortward of 
the line peak because of H {\sc i} absorption by gas clouds in the system
and intergalactic H {\sc i} gas (e.g., Hu et al. 2002; Ajiki et al. 2002).
Another important spectral feature of LAEs is that 
there is little continuum emission shortward of the Ly$\alpha$ line
(Hu et al. 2002; Kodaira et al. 2003). 
Such a continuum break is actually
seen in our two objects, SDF J132415.7+273058 and SDF J132418.3+271455,
which are already identified as LAEs at $z \approx 6.6$ 
(Kodaira et al. 2003). However, the other objects show no continuum feature
at wavelengths longer than the line peak, and thus we cannot identify LAEs
solely by using the continuum-break feature.

Among the remaining emission-line objects, the following two have a narrow line width:
6.5 \AA ~ for SDF J132518.8+273043, and 5.5 \AA ~ for SDF J132522.3+273520.
If the emission line were [O {\sc ii}] emission, the redshift
would be $z \approx 1.47$. Since the [O {\sc ii]} feature is a doublet
line of [O {\sc ii}]$\lambda$3726.0 and [O {\sc ii}]$\lambda$3728.8,
the line separation would be wider than 6.9 \AA\  at $z=1.47$ and the two
components could be resolved in our Echelle spectroscopy.
Further, if the line were H$\beta$, [O {\sc iii}]$\lambda$4959,
[O {\sc iii}]$\lambda$5007, or H$\alpha$ line, we would detect
some other emission lines in our spectra. Therefore, these two narrow-line 
objects must be a LAE (e.g., Taniguchi et al. 2003a).
 
As for the other ten emission-line objects, we found that the following 
five objects show an asymmetric profile with a sharp cutoff at wavelengths 
shortward of the line peak:
SDF J132352.7+271622, SDF J132353.1+271631, SDF J132408.3+271543,
SDF J132418.4+273345, and SDF J132432.5+271647. 
Therefore, the nine objects discussed above appear to be nice
candidates for LAEs at $z \approx$ 6.5 -- 6.6.
The remaining five emission-line objects show either
a symmetric emission-line profile or poor-S/N spectra.

In order to secure our spectral classification, we estimated the flux
ratio between $f_{\rm red}$ and $f_{\rm blue}$, where $f_{\rm red}$ 
is the flux at wavelengths longer than the emission-line peak, 
while $f_{\rm blue}$ is that at wavelengths longer than the emission-line peak. 
In table 4, we give our results. In figure 6, we show the frequency 
distribution of log $f_{\rm red}$/$f_{\rm blue}$ for the sample of 
14 emission-line objects. In this figure, we find bi-modal distributions
in log $f_{\rm red}$/$f_{\rm blue}$; i.e., the nine objects thought
to be nice LAE candidates have $f_{\rm red}$/$f_{\rm blue} > 1$,
while the remaining five objects have $f_{\rm red}$/$f_{\rm blue} <1$.
Some other observational tests have been proposed to identify
LAEs at high redshift (e.g., Rhoads et al. 2003; Kurk et al. 2004);
the S/N in our optical spectra prevents us from applying such tests.

In total, 
we identified nine LAEs whose emission-line shapes show a sharp cutoff at
wavelengths shortward of the line peak. They are listed up in the upper part
of table 2. Their optical spectra are shown in figure 7.



Since the remaining five emission-line objects show either 
a symmetric emission-line profile, or poor-S/N spectra,
we did not include them as our final LAE sample,
although we cannot rule out the possibility that they are LAEs at $z \approx 6.6$.
They are listed in the middle part of table 2, with a Ly$\alpha$ redshift, 
if they are a LAE. Their spectra are shown in figure 8.

We found no emission-line feature in the six candidates in our spectroscopic
sample (Nos. 10, 11, 13, 14, 15, and 20 in table 2). 
The estimated noises in our FOCAS spectra in these six objects
range from $2 \times 10^{-19}$ erg s$^{-1}$ cm$^2$ \AA$^{-1}$
to $3 \times 10^{-19}$ erg s$^{-1}$ cm$^2$ \AA$^{-1}$.
Taking account of these noises, we could detect 
their Ly$\alpha$ fluxes, expected from  our NB921 imaging, 
if they had an unresolved Ly$\alpha$ emission line in our FOCAS Echelle spectra
(6.4 \AA ~ resolution), because  their peak fluxes would range from
$3 \times 10^{-19}$ erg s$^{-1}$ cm$^2$ \AA$^{-1}$ to
$6 \times 10^{-19}$ erg s$^{-1}$ cm$^2$ \AA$^{-1}$.
If the line had some broad
feature, or if the line was recorded in some noisy spectral parts,
it seems that it would have been 
quite difficult to detect unambiguous emission-line
feature for these six objects in our FOCAS spectroscopy.
Since there was no emission-line feature in these six objects,
we cannot judge whether they are LAEs or low-$z$ emission-line sources.
Another possibility is that they might be
LBGs with weak Ly$\alpha$ emission. In any case, we need future very deep
optical spectroscopy for these objects.


Thumbnail images of the nine LAEs are shown in figure 9;
their photometric properties are given in table 2.
The redshift given in the last column of table 2 was estimated from the peak
of the Ly$\alpha$ emission line. The measurement error of redshift was
estimated to be $\pm$ 0.002.
The observed flux of Ly$\alpha$ emission of each object is given in
table 5. The Ly$\alpha$ line width is also given in this table.
The line width ranges from 180 km s$^{-1}$ to 460 km s$^{-1}$,
being comparable to those of previously identified LAEs beyond $z=5$.
Our spectra show no evidence of N {\sc v} $\lambda$ 1240
emission for all the nine LAEs.
These properties suggest that they are star-forming galaxies, rather than
active galactic nuclei.
LAEs with FWHM $> 300$ km s$^{-1}$ (four among the nine) may
experience a superwind in some cases (e.g., Dawson et al. 2001;
Ajiki et al. 2002).


\subsection{A Fraction of LAEs}

Our optical spectroscopy has confirmed that 9 among 20 candidates
are real LAEs at $z \approx 6.6$. 
Five among the remaining 11 objects are single-line emitters with 
a symmetric emission-line profile. They may be either a LAE at 
$z \sim 6.6$ or an [O {\sc ii}] $\lambda$ 3727 emitter at $z \sim 1.47$.

Since it is safe to keep the five single-line objects as unclassified 
objects, we obtain from our spectroscopy that the fraction of
reliable LAEs in our photometric sample is $f$(LAE) = 9/20 = 0.45.
This can be regarded as a lower limit in our study.
If we include nominally 5 single-line emitters into the LAE sample,
we obtain  $f$(LAE) $= 14/20 = 0.70$. The real fraction of LAEs
in our study seems to be in the range of  $f$(LAE) $\simeq 0.45$ -- 0.70.

\subsection{Morphological Structures of LAEs}

Since our final PSF size is 0.$''$98, it seems difficult to
investigate the structural properties of the nine LAEs. However,
some NB921 CCD frames were obtained under 0.$''$5 - 0.$''$7 conditions
during our observing runs. Using only these data, we made
a high-resolution NB921 image with a PSF of 0.$''$71. The integration
time for this image was 37215.7 seconds, $\simeq$ 70\% of the total
integration time (see table 1).

In figure 10, we show newly made NB921 images of the nine LAEs.
Their azimuthal radial profiles are also shown in figure 11. 
Their sizes (FWHM) range from 0.81 arcsec to 1.02 arcsec,
as given in table 6. Although we do not conclude that all LAEs
are spatially resolved in our NB921 images, some of them
(e.g., Nos. 4 and 8) appear to be spatially extended.
After deconvolution with the PSF size, we obtain their sizes
between 0.39 and 0.73 arcsec, corresponding to $\simeq$
(2 -- 4) $h_{0.7}^{-1}$ kpc.
It is thus suggested that gaseous matter around LAE host galaxies
are spatially extended up to several kpc at $z \approx 6.6$ in some cases.
Imaging with Advanced Camera for Surveys on board the Hubble Space Telescope
with the $z$(F850LP) filter will be important to investigate their
detailed Ly$\alpha$ morphologies. 



\subsection{Comments on Possible Amplification by Gravitational Lensing}

The SDF is a so-called blank field
and thus there is no apparent cluster of galaxies  known to date
at low and intermediate redshifts in our field.
In this respect, our survey may not suffer from strong
amplification by gravitational lensing, unlike several surveys
made with the help of gravitational lensing (Ellis et al. 2001;
Hu et al. 2002; Santos et al. 2004; Kneib et al. 2004; Pell\'o et al. 2004a).
However, any high-$z$ objects could suffer from gravitational lensing
because of a larger optical depth for lensing (Wyithe, Loeb 2002;
Shioya et al. 2002). Therefore,
it is possible that some LAEs found in our survey
could be gravitationally amplified by a foreground galaxy lying
close to their lines of sight.

In order to check such possibilities,
we examined carefully our $i^\prime$ and $z^\prime$ images
for all the nine objects. We also examined our $B$, $V$, and $R$ images
to check whether or not there are any foreground galaxies.
As shown in figures 9 and 10, there are some galaxies that could be foreground
galaxies around each LAE. For example,
a probable foreground object is found to be located at 0.$''$8 NW of
SDF J132418.3+271455. We listed all such foreground galaxies within
a radius of 4$''$.
We then evaluated the magnification factor for each LAE in the following way
(Shioya et al. 2002; see also Yamada et al. 2003):
(1) We estimate its photometric redshift and rest-frame $B$-band
luminosity based on our optical broad-band photometric data for
each foreground galaxy, if any.
(2) We estimated its stellar velocity dispersion using
the Tully--Fisher relation and the $B$-band luminosity estimated above.
(3) We estimated the magnification factor based on the singular
isothermal sphere model for gravitational lensing.
We found that the amplification
factor by gravitational lensing is smaller than 1.1 for all the cases.
It is also noted that there is no counter image for each LAE within 
the limiting magnitude, and thus
the gravitational amplification factor should be less than a factor of $\sim2$,
being consistent with the above factor.
Therefore, we conclude that our LAE sample did  not suffer from
strong amplification by the gravitational lensing.
This allows us to perform simple statistical analyses for our sample of LAEs,
such as the number density and the star-formation rate density at $z=6.6$.

\section{DISCUSSION}

\subsection{The Number Density of LAEs at $z \approx 6.6$}

First, we estimated the number density of LAEs at $z\simeq 6.6$ 
and compared it with those at other redshifts. 
Our sample of LAE candidates consists of 58 objects.
Then, our optical spectroscopy found that nine objects are 
LAEs at $z\simeq 6.5$ -- 6.6 among the 20 objects in our spectroscopic sample.
Therefore, from a statistical point of view, $\approx 45\%$ 
of our photometrically selected candidates are expected to be
LAEs at $z\approx 6.6$; 
note again that the nine galaxies for which optical spectroscopy was 
made were randomly selected from the photometric sample. 
The limiting magnitude of our sample, $NB921=26.54$, 
corresponds roughly to 
$f({\rm Ly}\alpha)\simeq 4.1 \times 10^{-18}$ erg s$^{-1}$ cm$^{-2}$,
or $L({\rm Ly}\alpha) \simeq 2.0\times 10^{42}$ 
$h_{0.7}^{-2}$ erg s$^{-1}$ for our adopted cosmology.
Since the survey volume of our survey is 217200 $h_{0.7}^{-3}$ 
Mpc$^3$, we obtain the number density of LAEs brighter than 
this limiting luminosity to be $n({\rm LAE}) \simeq 
58 \times 0.45/217200 \simeq 1.2 \times 10^{-4}$ 
$h_{0.7}^3$ Mpc$^{-3}$.

It is interesting to compare our result with previous studies
on LAEs at high redshift. Note, however, that the detection 
completeness is not taken into account in later discussion
because it is difficult to estimate it for each survey.
Ouchi et al. (2003) have made a wide-field survey of LAEs 
at $z\simeq 4.9$. 
From their data, we estimate that the number density of LAEs 
at $z\simeq 4.9$ brighter than 
$L({\rm Ly}\alpha) = 2.4\times 10^{42}$ $h_{0.7}^{-2}$ erg s$^{-1}$ 
is $\simeq 2 \times 10^{-4}$ $h_{0.7}^3$ Mpc$^{-3}$
(M. Ouchi, private communication), 
which is close to that for our $z \simeq 6.6$ LAEs.
Rhoads and Malhotra (2001) also carried out a similar survey 
for $z=5.7$ LAEs; the limiting luminosity of their survey, 
$\approx 5.3 \times 10^{42}$ $h_{0.7}^{-2}$ erg s$^{-1}$, 
is brighter than ours (see also Rhoads et al. 2003) .
They found in their sample the number density of LAEs 
to be $\simeq 4\times 10^{-5}$ $h_{0.7}^3$ Mpc$^{-3}$.
If we limit our $z\simeq 6.6$ LAE sample to
Rhoads et al.'s (2003) limiting luminosity, 
we obtain $\simeq 2 \times 10^{-5}$ $h_{0.7}^3$ Mpc$^{-3}$.
Similarly, we obtain $\simeq 4 \times 10^{-5}$ $h_{0.7}^3$ Mpc$^{-3}$ 
for Ouchi et al.'s (2003) sample.
These results indicate that the number density of LAEs 
does not change significantly from $z\simeq 4.9$ to $6.6$.
Combined with a result given in Ouchi et al. (2003) 
that the number density of $z \simeq 4.9$ LAEs 
is not clearly different 
from that of $z=3.4$ LAEs in Cowie and Hu's (1998) sample, 
we conclude that LAEs do not evolve in number
from $z\simeq 3.4$ to $6.6$ within the uncertainties in the data.

\subsection{The Star-Formation Rate of LAEs at $z \approx 6.6$}

\subsubsection{The Star-Formation Rate Estimated from Ly$\alpha$ Luminosity}

Next, we estimate the star-formation rate of the LAEs at $z \approx 6.6$.
We have two kinds of SFRs: one is based on our optical spectroscopy and
the other is based on our continuum-subtracted NB921 data.

First, we discuss the SFRs based on our spectroscopy.
The observed Ly$\alpha$ flux, Ly$\alpha$ luminosity,
and star-formation rate, SFR(Ly$\alpha$), of each LAE are summarized 
in table 5.
Note that the SFR(Ly$\alpha$) is estimated by using 
the relation (Kennicutt 1998; Brocklehurst 1971)

\begin{equation}
SFR({\rm Ly}\alpha)  = 9.1 \times 10^{-43} L({\rm Ly}\alpha) ~ 
M_\odot ~ {\rm yr}^{-1},
\end{equation}
where the Salpeter initial mass function with ($m_{\rm lower}$, $m_{\rm upper}$)
= (0.1 $M_\odot$, 100 $M_\odot$) is adopted.
The SFRs obtained for the nine LAEs range from 
$\approx 3$ $h_{0.7}^{-2}$ $M_\odot$ yr$^{-1}$
to $\approx 9$ $h_{0.7}^{-2}$ $M_\odot$ yr$^{-1}$ with an average of 5.7 $\pm$ 2.3
$h_{0.7}^{-2}$ $M_\odot$ yr$^{-1}$. These values are comparable to those of LAEs at 
$z \simeq$ 5.1 -- 5.8 (e.g., Taniguchi et al. 2003b and references therein).
The total spectroscopic SFR for the nine LAEs amounts to 

\begin{equation}
SFR^{\rm spec}({\rm Ly}\alpha) \simeq 49.4 ~ h_{0.7}^{-2} ~ M_\odot ~ {\rm yr}^{-1}.
\end{equation}
It should be mentioned that the SFRs estimated above are lower limits
because it is quite likely that a blue half or more of the Ly$\alpha$ emission
may be absorbed by dust grains in the galaxy, itself, and by
the intergalactic H {\sc i} gas (e.g., Hu et al. 2002; Haiman 2002).
Therefore, it would be desirable to study the rest-frame UV-optical continuum
(Hu et al. 2002; Kodaira et al. 2003).
 
Second, we discuss SFRs based on the continuum-subtracted NB921 data;
$SFR^{\rm image}$(Ly$\alpha$).
Our results are summarized in table 7; note that we give the 
photometric SFRs  for all 58 LAE candidates. 
The frequency distributions of Ly$\alpha$ luminosity
for all 58 LAE candidates are shown in figure 12.
The Ly$\alpha$ luminosity of the two
sources, SDF J132353.1+271631 (No. 2) and SDF J132518.8+273043, is given as
an upper-limit. This is due to the fact that the bandpass-corrected $z^\prime$ flux
($f_{9500}$; see table 2) is estimated to be a bit higher because of 
low signal-to-noise ratios.
Although these two upper limit data are present, 
the SFRs of the nine LAEs confirmed
by our spectroscopy range between 3 and 10 $h_{0.7}^{-2}$ $M_\odot {\rm yr}^{-1}$,
being similar to that obtained in our spectroscopy. 
The total photometric SFR for the nine LAEs is estimated to be 

\begin{equation}
SFR^{\rm image}({\rm Ly}\alpha) \simeq 58.5 ~ h_{0.7}^{-2} ~ M_\odot ~ {\rm yr}^{-1},
\end{equation}
being nearly the same as that obtained by the spectroscopy.
In figure 13, we compare the observed fluxes of Ly$\alpha$ emission obtained from
spectroscopy and imaging for the nine LAEs.



\subsubsection{The Star Formation Rate Estimated from UV Luminosity}

We now estimate another SFR derived from the UV continuum luminosity
for our sample. The observed $z^\prime$ magnitude can be converted to
a UV continuum luminosity at $\lambda = 1260$ \AA.
Using the relation (Kennicutt 1998; see also Madau et al. 1998),

\begin{equation}
\label{UVtoSFR}
SFR({\rm UV}) = 1.4 \times 10^{-28}L_{\nu}({\rm UV})
~~ M_{\odot} ~ {\rm yr}^{-1}
\end{equation}
where $L_\nu$(UV) is the UV continuum luminosity
in units of erg s$^{-1}$ Hz$^{-1}$,
we estimate the SFR based on the rest-frame UV ($\lambda=1260$\AA)
continuum luminosity for each object. The results are summarized
in table 7. Note that the same initial mass function for $SFR$(Ly$\alpha$)
was adopted in this estimate. 
The frequency distributions of UV continuum luminosity 
for all 58 LAE candidates are shown in figure 14.


\subsubsection{Comparisons between SFR(Ly$\alpha$) and SFR(UV)}

We then compared the SFRs derived from the Ly$\alpha$ and the UV photometric
data. The result is shown in figure 15; also see table 7.
As expected from previous studies (e.g., Hu et al. 2003; Kodaira et al. 2003;
Ajiki et al. 2003), we find that $SFR$(UV) tends to be
higher by a factor of 5, on average, than $SFR^{\rm image}$(Ly$\alpha$).
The obtained factor seems to be higher than the previous results;
e.g., by a factor of 2.


\subsection{The Star Formation Rate Density of LAEs at $z \approx 6.6$}

We now discuss the star-formation history beyond
redshift 6 based on our data. 
We can estimate the total star-formation rate of 58 LAEs in our 
photometric sample using the equivalent width of the NB921 flux; 
$SFR^{\rm image}$(total) $\simeq 276$ $h_{0.7}^{-2}$  $M_\odot$ yr$^{-1}$.
Given the survey volume, 217200 $h_{0.7}^{-3}$ Mpc$^{3}$, we nominally obtain
a star-formation rate density (SFRD) of
$\rho_{\rm SFR, upper} \simeq 1.3 \times 10^{-3}$
$h_{0.7}$ $M_\odot$ yr$^{-1}$  Mpc$^{-3}$.
However, this value should be regarded being as an upper bound obtained from
our study. Since we estimated the fraction of LAEs in our photometric sample,
$f$(LAE) $\simeq 0.45$, it seems reasonable to assume that
45\% of the LAE candidates are at $z \approx 6.6$. We then obtained 
the corrected total SFR,  $SFR^{\rm image, cor}({\rm total})
\simeq 276 \times 0.45 \simeq 124$ $h_{0.7}^{-2}$ $M_\odot$ yr$^{-1}$.
This gives $\rho_{\rm SFR} \simeq 5.7 \times 10^{-4}$
$h_{0.7}$ $M_\odot$ yr$^{-1}$  Mpc$^{-3}$, being consistent with our
previous study (Kodaira et al. 2003). Recently, Kurk et al. (2004) made
a grism survey for LAEs at $z \approx 6.4$ -- 6.6 and then identified
a LAE at $z=6.518$. Their new type of survey also gave a similar value of SFRD:
$\rho_{\rm SFR} \simeq 5.0 \times 10^{-4}$ $h_{0.7}$ $M_\odot$ yr$^{-1}$  Mpc$^{-3}$. 

We were also able to obtain a lower bound of the total SFR using the observed Ly$\alpha$
fluxes obtained with our spectroscopy. Using the data given in table 5, we
obtained $SFR^{\rm spec}$(total) $\simeq 49.4$ $h_{0.7}^{-2}$ $ M_\odot$ yr$^{-1}$.
This gives a lower bound of SFRD, $\rho_{\rm SFR, lower}
\simeq 2.3 \times 10^{-4}$ $h_{0.7}$ $M_\odot$ yr$^{-1}$  Mpc$^{-3}$.

It is interesting to compare this value with previous estimates
between $z = 0$ and $z \sim 6$.
In figure 16, we compare this star-formation rate density with those
of previous studies compiled by Ajiki et al (2003) under the
same cosmological parameters as those adopted in this paper. In addition,
we also show the results obtained from LAE surveys by Cowie and Hu (1998),
Kudritzki et al. (2000), Ouchi et al. (2003), and Ajiki et al. (2003)
and from LBG surveys by Giavalisco et al. (2004b) and Dickinson et al. (2004).


Our SFRD together with those based on the LAE surveys tend to give
smaller values with respect to those obtained from LBG ones at 
$z \sim$ 3 -- 6. 
It should be reminded here again that
we applied neither any reddening correction nor integration by assuming
a certain luminosity function for the LAEs. 
Previous studies have shown that the SFR based on the Ly$\alpha$ luminosity
is smaller by a factor of two or more than that based on the UV continuum
(Hu et al. 2002; Kodaira et al. 2003). Kodaira et al. (2003) also showed that
the SFR based on the rest-frame optical continuum gives a value several times
as high as that based on the Ly$\alpha$ luminosity.
More recently, Reddy and Steidel (2004) have shown from an analysis of 
GOODS multiwavelength data (see for GOODS, Giavalisco et al. 2004a) that
the SFR based on X-ray data is higher by a factor of 5 than that based
on the UV luminosity. 
In addition, integration using a luminosity function of LAEs also
increases the SFR by a factor of two or more (see Ajiki et al. 2003).
Although we do not still have a well-defined Ly$\alpha$ luminosity function
based on a large sample of LAEs (e.g., Santos et al. 2004; Hu et al. 2004),
the SFR derived here for our sample of LAEs at $z \approx$ 6.6 may be
smaller by a factor of several than the real value; i.e.,  $\rho_{\rm SFR}$
is on the order of $10^{-2}$ $h_{0.7}^{-2}$ $M_\odot$ yr$^{-1}$ Mpc$^{-3}$,
 or higher.

Finally, we comment on SFRDs based on so-called $i^\prime$-dropout samples.
In addition to the follow-up optical spectroscopy of LAE candidates 
at $z \approx 6.6$ (this work), we also conducted optical spectroscopy
of a sample of
$i^\prime$-dropouts found in the SDF (details will be given elsewhere). 
During this follow-up spectroscopy, we found a LAE at $z=6.33$ (Nagao et al. 2004).
Its $z^\prime$-band flux is dominated by strong Ly$\alpha$ emission, rather than
the stellar UV continuum. If such strong LAEs could be involved in a photometric
sample of $i^\prime$-dropouts, we would overestimate SFRD. This issue
shall be addressed in our forthcoming papers.

We would like to thank  the Subaru Telescope staff for
their invaluable assistance. We would also like to thank an
anonymous referee for his/her careful reading the paper and
many useful suggestions and comments.
This work was financially supported in part by
the Ministry of Education, Culture, Sports, Science and Technology
(Nos. 10044052, and 10304013) and JSPS (No. 15340059).
MA and TN are JSPS fellows.


\begin{table}
  \caption{Journal of optical imaging observations.}\label{tab:first}
  \begin{center}
    \begin{tabular}{lccccc}
\hline
\hline
 Band & Exposure Time & $m_{\rm AB}$(lim)\footnotemark[*] \\
\hline
$B$     & 35700 & 28.45 \\
$V$     & 20400 & 27.74 \\
$R$     & 36000 & 27.80 \\
$i'$    & 48060 & 27.43 \\
$z'$    & 30240 & 26.62 \\
$NB816$ & 36000 & 26.63 \\
$NB921$ & 53940 & 26.54 \\
\hline
\multicolumn{3}{l}{
\footnotemark[*]{Limiting magnitude (AB) for a 3$\sigma$ detection
                   on a $2.''0$ diameter aperture.}
}
    \end{tabular}
  \end{center}
\end{table}


{\scriptsize
\begin{longtable}{cccccccc}
  \caption{Photometric properties of the LAEs\footnotemark[*].}\label{tab:second}
  \hline\hline
{No.} & {Name\footnotemark[$\dagger$]} & 
\multicolumn{4}{c}{Optical AB Magnitude\footnotemark{$\ddagger$}} &
{$EW_{\rm obs}$}& {$z$}\\
{} & {} & {$i^\prime$} & {$z^\prime$} & {$NB921$} & {$z_{9500}$} & {(\AA)} & {} \\
\endfirsthead
\endhead
  \hline
\endfoot
  \hline
\endlastfoot
\hline
 1 & SDF J132352.7+271622 & $>$28.62 & $>$27.81 & 25.68 & $>$26.42 & $>$271 & 6.542  \\
 2 & SDF J132353.1+271631 & $>$28.62 &   26.62  & 25.21 &   25.57  &    132 & 6.540  \\
 3 & SDF J132408.3+271543 &   27.56  &   25.96  & 24.49 &   24.94  &    147 & 6.554  \\
 4 & SDF J132415.7+273058\footnotemark{$\S$}
                          &   27.33  &   25.69  & 24.13 &   24.73  &    181 & 6.541  \\
 5 & SDF J132418.3+271455\footnotemark{$\S$}
                          &  (28.40) & $>$27.81 & 24.98 & $>$26.42 & $>$588 & 6.578  \\
 6 & SDF J132418.4+273345 & $>$28.62 &  (27.34) & 25.17 & $>$26.42 & $>$408 & 6.506  \\
 7 & SDF J132432.5+271647 & $>$28.62 & $>$27.81 & 25.76 & $>$26.42 & $>$298 & 6.580  \\
 8 & SDF J132518.8+273043 &  (28.42) &   26.81  & 25.31 &  (25.81) &    158 & 6.578  \\
 9 & SDF J132522.3+273520 &  (28.21) &   26.49  & 24.87 &   25.57  &    204 & 6.597  \\
\hline
10 & SDF J132406.5+271634 & $>$28.62 &   26.69  & 25.41 &   25.58  &     98 &  \\
11 & SDF J132410.8+271928 & $>$28.62 &   26.27  & 25.08 &   25.13  &     81 &  \\
12 & SDF J132417.9+271746 & $>$28.62 & $>$27.81 & 25.44 & $>$26.42 & $>$338 & (6.55) \\
13 & SDF J132428.7+273049 & $>$28.62 & $>$27.81 & 25.63 & $>$26.42 & $>$289 &  \\
14 & SDF J132500.9+272030 & $>$28.62 &  (27.40) & 25.57 & $>$26.42 & $>$256 &  \\
15 & SDF J132515.5+273714 & $>$28.62 & $>$27.81 & 25.43 & $>$26.42 & $>$342 &  \\
16 & SDF J132518.4+272122 & $>$28.62 &   26.40  & 24.24 &  (26.12) &    732 & (6.55) \\
17 & SDF J132519.9+273704 & $>$28.62 &  (27.20) & 25.34 & $>$26.42 & $>$318 & (6.54) \\
18 & SDF J132520.4+273459 & $>$28.62 &   26.89  & 25.40 &  (26.88) &    154 & (6.60) \\
19 & SDF J132521.1+272712 & $>$28.62 &   26.66  & 25.16 &   25.65  &    157 & (6.55) \\
20 & SDF J132525.3+271932 & $>$28.62 &   26.70  & 25.60 &   25.52  &     65 &  \\
\hline
21 & SDF J132338.4+274652 & $>$28.62 & $>$27.81 & 25.85 & $>$26.42 & $>$278 &  \\
22 & SDF J132338.6+272940 &  (28.40) &   26.68  & 25.81 & $>$26.42 & $>$258 &  \\
23 & SDF J132342.2+272644 & $>$28.62 & $>$27.81 & 25.45 & $>$26.42 & $>$337 &  \\
24 & SDF J132343.2+272452 & $>$28.62 &   26.68  & 24.59 &  (26.25) &    581 &  \\
25 & SDF J132347.7+272360 & $>$28.62 &  (27.66) & 25.55 & $>$26.42 & $>$283 &  \\
26 & SDF J132348.9+271530 & $>$28.62 & $>$27.81 & 25.59 & $>$26.42 & $>$309 &  \\
27 & SDF J132349.2+273211 &  (28.50) & $>$27.81 & 25.79 & $>$26.42 & $>$226 &  \\
28 & SDF J132349.2+274546 & $>$28.62 &    27.77 & 25.20 & $>$26.42 & $>$429 &  \\
29 & SDF J132353.4+272602 &  (28.10) &   26.66  & 24.97 &  (25.79) &    236 &  \\
30 & SDF J132357.1+272448 & $>$28.62 &  (27.16) & 25.63 &  (25.17) &    165 &  \\
31 & SDF J132401.5+273837 &  (28.53) &  (27.71) & 25.76 & $>$26.42 & $>$223 &  \\
32 & SDF J132402.6+274653 & $>$28.62 &   26.70  & 25.46 &   25.57  &     90 &  \\
33 & SDF J132410.5+272811 & $>$28.62 & $>$27.81 & 25.91 & $>$26.42 & $>$250 &  \\
34 & SDF J132419.3+274125 & $>$28.62 & $>$27.81 & 25.56 & $>$26.42 & $>$363 &  \\
35 & SDF J132422.6+274459 & $>$28.62 &   26.91  & 25.24 &  (26.02) &    223 &  \\
36 & SDF J132424.2+272649 &  (28.53) &  (27.14) & 25.77 &  (26.07) &    120 &  \\
37 & SDF J132425.4+272410 & $>$28.62 &  (27.58) & 25.57 & $>$26.42 & $>$199 &  \\
38 & SDF J132425.9+274324 & $>$28.62 & $>$27.81 & 25.56 & $>$26.42 & $>$271 &  \\
39 & SDF J132425.0+273606 &  (28.56) &  (27.43) & 25.76 & $>$26.42 & $>$292 &  \\
40 & SDF J132434.3+274056 & $>$28.62 & $>$27.81 & 25.71 & $>$26.42 & $>$315 &  \\
41 & SDF J132435.0+273957 & $>$28.62 & $>$27.81 & 25.45 & $>$26.42 & $>$339 &  \\
42 & SDF J132436.6+272223 & $>$28.62 & $>$27.81 & 25.90 & $>$26.42 & $>$229 &  \\
43 & SDF J132440.2+272553 & $>$28.62 & $>$27.81 & 25.83 & $>$26.42 & $>$284 &  \\
44 & SDF J132443.4+272633 & $>$28.62 &   26.88  & 25.37 &  (25.89) &    163 &  \\
45 & SDF J132444.4+273942 & $>$28.62 &  (27.29) & 25.41 & $>$26.42 & $>$299 &  \\
46 & SDF J132445.6+273033 &  (28.20) &   26.97  & 25.20 &  (26.16) &    276 &  \\
47 & SDF J132447.7+271106 & $>$28.62 &   26.92  & 25.37 &  (25.95) &    174 &  \\
48 & SDF J132449.5+274237 & $>$28.62 &  (27.46) & 25.88 & $>$26.42 & $>$171 &  \\
49 & SDF J132450.7+272160 & $>$28.62 & $>$27.81 & 25.60 & $>$26.42 & $>$289 &  \\
50 & SDF J132455.4+271314 & $>$28.62 &  (27.76) & 25.67 & $>$26.42 & $>$252 &  \\
51 & SDF J132455.8+274015 & $>$28.62 &  (27.40) & 25.75 & $>$26.42 & $>$198 &  \\
52 & SDF J132458.5+273913 & $>$28.62 &  (27.60) & 25.56 & $>$26.42 & $>$650 &  \\
53 & SDF J132458.0+272349 & $>$28.62 &  (27.08) & 24.71 & $>$26.42 & $>$273 &  \\
54 & SDF J132503.4+273838 &  (28.23) & $>$27.81 & 25.94 & $>$26.42 & $>$207 &  \\
55 & SDF J132506.4+274047 & $>$28.62 & $>$27.81 & 25.95 & $>$26.42 & $>$193 &  \\
56 & SDF J132516.7+272236 & $>$28.62 &  (27.54) & 25.36 & $>$26.42 & $>$341 &  \\
57 & SDF J132528.0+271328 & $>$28.62 &  (27.76) & 25.71 & $>$26.42 & $>$242 &  \\
58 & SDF J132533.4+271420 & $>$28.62 & $>$27.81 & 25.65 & $>$26.42 & $>$336 &  \\
\hline
\multicolumn{8}{l}{
\footnotemark[*]{The LAEs are not detected in $B$, $V$, and $R$; see, for
                  3$\sigma$ upper limits in these bands, table 1.}
}\\
\multicolumn{8}{l}{
\footnotemark[$\dagger$]{The sky position, $\alpha$(J2000) and $\delta$(J2000),
                  is given in the name.}
}\\
\multicolumn{8}{l}{
\footnotemark[$\ddagger$]{AB magnitude in a $2^{\prime\prime}$ diameter.
                  The magnitudes between the 1 $\sigma$ and 2 $\sigma$ detection
                  levels are put in parentheses.}
}\\
\multicolumn{8}{l}{
\footnotemark[$\S$]{LAE identified in Kodaira et al. (2003).}
}\\
\end{longtable}
}


\begin{table}
  \caption{Journal of optical spectroscopy.}\label{tab:thid}
  \begin{center}
    \begin{tabular}{ccccl}
\hline
\hline
{MOS ID} & {Setting\footnotemark[*]} & {Exposure Time} & {seeing} & 
{Object ID\footnotemark[$\dagger$]} \\
{} & {} & {(s)} & {($''$)} & {} \\
\hline
Mask 1 & 300R+O58           & 19800 & 0.5--0.8 & 5, 7 \\
Mask 2 & 300R+O58           & 10800 & 0.7--0.8 & 4, 6 \\
Mask 4 & Echelle+$z^\prime$ & 14400 & 0.6--0.7 & 1, 2, 3 \\
Mask 5 & 300R+O58           & 12600 & 0.6--1.0 & 8 \\
Mask 6 & Echelle+$z^\prime$ & 13500 & 0.5-0.7 & 9 \\
\hline
\multicolumn{5}{l}{
\footnotemark[*]{Grating + order-cut filter.}
}\\
\multicolumn{5}{l}{
\footnotemark[$\dagger$]{Object ID numbers that correspond to the ones
given in tables 2, 4, 5, 6 and 7.}
}
    \end{tabular}
  \end{center}
\end{table}


\begin{table}
  \caption{Spectral identification of emission-line objects.}\label{tab:fourth}
  \begin{center}
    \begin{tabular}{ccc}
\hline
\hline
{ID} & {$f_{\rm red}/f_{\rm blue}$} & {log $f_{\rm red}/f_{\rm blue}$} \\
\hline
 1  &  3.60$\pm$2.08 &  0.556$\pm$0.251 \\
 2  &  1.22$\pm$0.43 &  0.086$\pm$0.153 \\
 3  &  1.48$\pm$0.32 &  0.170$\pm$0.094 \\
 4  &  1.33$\pm$0.41 &  0.124$\pm$0.134 \\
 5  &  1.29$\pm$0.40 &  0.111$\pm$0.135 \\
 6  &  1.33$\pm$0.18 &  0.124$\pm$0.059 \\
 7  &  1.41$\pm$0.46 &  0.149$\pm$0.142 \\
 8  &  1.15$\pm$0.54 &  0.061$\pm$0.204 \\
 9  &  1.78$\pm$0.43 &  0.250$\pm$0.105 \\
\hline
12  &  0.63$\pm$0.47 & $-$0.201$\pm$0.324 \\
16  &  0.47$\pm$0.62 & $-$0.328$\pm$0.573 \\
17  &  0.86$\pm$0.44 & $-$0.151$\pm$0.222 \\
18  &  0.86$\pm$0.78 & $-$0.151$\pm$0.394 \\
19  &  0.54$\pm$0.57 & $-$0.268$\pm$0.459 \\
\hline
    \end{tabular}
  \end{center}
\end{table}


\begin{table}
  \caption{Star-formation properties of the nine LAEs.}\label{tab:fifth}
  \begin{center}
    \begin{tabular}{ccccccc}
\hline
\hline
{No.} & {Name} & {$f^{\rm spec}$(Ly$\alpha$)} & {$L^{\rm spec}$(Ly$\alpha$)} &
{$SFR^{\rm spec}$(Ly$\alpha$)} & \multicolumn{2}{c}{FWHM} \\ 
{} & {} & {($10^{-18}$ erg s$^{-1}$ cm$^{-2}$)} &
{($10^{42}$ $h_{0.7}^{-2}$ erg s$^{-1}$)} & {($h_{0.7}^{-2}$ $M_\odot$ yr$^{-1}$)} &
{(\AA)} & {(km s$^{-1}$)} \\
\hline
1 & SDF J132352.7+271622 & 7.31  &  3.56  & 3.24  & 14.0  & 458 \\
2 & SDF J132353.1+271631 & 8.22  &  4.00  & 3.64  & 10.6  & 347 \\
3 & SDF J132408.3+271543 & 16.60 &  8.13  & 7.39  &  7.5  & 245 \\
4 & SDF J132415.7+273058 & 19.60 &  9.55  & 8.69  & 10.2  & 334 \\
5 & SDF J132418.3+271455 & 10.95 &  5.41  & 4.92  & $<$ 9.0 & $<$ 293 \\
6 & SDF J132418.4+273345 & 19.43 &  9.35  & 8.51  &  7.8  & 256 \\
7 & SDF J132432.5+271647 &  6.17 &  3.05  & 2.77  &  9.7  & 316 \\
8 & SDF J132518.8+273043 &  7.29 &  3.60  & 3.28  &  6.5  & 212 \\
9 & SDF J132522.3+273520 & 15.48 &  7.69  & 7.00  &  5.5  & 179 \\
\hline
    \end{tabular}
  \end{center}
\end{table}


\begin{table}
  \caption{Full width at half maximum size of the nine LAEs.}\label{tab:sixth}
  \begin{center}
    \begin{tabular}{ccccc}
\hline
\hline
{No.} & {Name} & {$FWHM$} & {$FWHM_{\rm cor}$\footnotemark[*]} & {$FWHM_{\rm cor}$} \\
{} & {} & {($''$)} & {($''$)} & {($h_{0.7}^{-1}$ kpc)} \\
\hline
1 & SDF J132352.7+271622 & 0.81$\pm$0.04 &  0.39$\pm$0.04  & 2.1$\pm$0.2 \\
2 & SDF J132353.1+271631 & 0.95$\pm$0.03 &  0.63$\pm$0.03  & 3.5$\pm$0.1 \\
3 & SDF J132408.3+271543 & 0.92$\pm$ 0.02 &  0.59$\pm$0.02  & 3.2$\pm$0.1 \\ 
4 & SDF J132415.7+273058 & 1.02$\pm$0.02 &  0.73$\pm$0.02  & 4.0$\pm$0.1 \\
5 & SDF J132418.3+271455 & 0.84$\pm$0.02 &  0.45$\pm$0.02  & 2.5$\pm$0.1 \\
6 & SDF J132418.4+273345 & 0.84$\pm$0.04 &  0.45$\pm$0.04  & 2.5$\pm$0.2 \\ 
7 & SDF J132432.5+271647 & 0.93$\pm$0.04 &  0.60$\pm$0.04  & 3.3$\pm$0.2 \\
8 & SDF J132518.8+273043 & 1.02$\pm$0.03 &  0.73$\pm$0.03  & 4.0$\pm$0.2 \\
9 & SDF J132522.3+273520 & 0.94$\pm$0.03 &  0.62$\pm$0.03  & 3.4$\pm$0.2 \\
\hline
\multicolumn{5}{l}{
\footnotemark[*]{Deconvolved with a PSF size of 0.$''$71.}
}
    \end{tabular}
  \end{center}
\end{table}


{\tiny
\begin{longtable}{cccccc}
  \caption{Star-formation properties of our LAE candidates.}\label{tab:seventh}
  \hline\hline
{No.} & {Name} & {$L^{\rm image}$(Ly$\alpha$)\footnotemark[*]} &
{$L_{\nu}$(UV)} & {$SFR^{\rm image}$(Ly$\alpha$)} & {$SFR$(UV)} \\
\endfirsthead
{} & {} & {($10^{42}$ erg s$^{-1}$)} &
{($10^{29}$ erg s$^{-1}$ Hz$^{-1}$)} & {($M_\odot$ yr$^{-1}$)} & {($M_\odot$ yr$^{-1}$)}\\
\endhead
  \hline
\endfoot
  \hline
\endlastfoot
\hline
 1 & SDF J132352.7+271622 &  4.4 & $<$6.4 &  4.0 & $<$9.0 \\
 2 & SDF J132353.1+271631 &  4.6 &   13.9 &  4.2 &   19.4 \\
 3 & SDF J132408.3+271543 &  9.2 &   25.0 &  8.4 &   35.0 \\
 4 & SDF J132415.7+273058 & 13.7 &   30.1 & 12.4 &   42.2 \\
 5 & SDF J132418.3+271455 &  9.4 & $<$6.4 &  8.6 & $<$9.0 \\
 6 & SDF J132418.4+273345 &  6.6 & $<$6.4 &  6.0 & $<$9.0 \\
 7 & SDF J132432.5+271647 &  4.8 & $<$6.4 &  4.4 & $<$9.0 \\
 8 & SDF J132518.8+273043 &  4.4 &   11.2 &  4.0 &   15.6 \\
 9 & SDF J132522.3+273520 &  7.1 &   13.9 &  6.5 &   19.5 \\
\hline
10 & SDF J132406.5+271634 &  3.4 &   13.9 &  3.1 &   19.4 \\
11 & SDF J132410.8+271928 &  4.3 &   21.0 &  3.9 &   29.4 \\
12 & SDF J132417.9+271746 &  5.4 & $<$6.4 &  4.9 & $<$9.0 \\
13 & SDF J132428.7+273049 &  4.6 & $<$6.4 &  4.2 & $<$9.0 \\
14 & SDF J132500.9+272030 &  4.1 & $<$6.4 &  3.7 & $<$9.0 \\
15 & SDF J132515.5+273714 &  5.5 & $<$6.4 &  5.0 & $<$9.0 \\
16 & SDF J132518.4+272122 & 15.4 &    8.4 & 14.0 &   11.7 \\
17 & SDF J132519.9+273704 &  5.1 & $<$6.4 &  4.7 & $<$9.0 \\
18 & SDF J132520.4+273459 &  4.0 &   10.5 &  3.7 &   14.6 \\
19 & SDF J132521.1+272712 &  5.1 &   12.9 &  4.6 &   18.1 \\
20 & SDF J132525.3+271932 &  2.4 &   14.6 &  2.2 &   20.4 \\
\hline
21 & SDF J132338.4+274652 &  4.5 & $<$6.4 &  4.1 & $<$9.0 \\
22 & SDF J132338.6+272940 &  4.1 & $<$6.4 &  3.8 & $<$9.0 \\
23 & SDF J132342.2+272644 &  5.4 & $<$6.4 &  4.9 & $<$9.0 \\
24 & SDF J132343.2+272452 & 10.9 &    7.5 &  9.9 &   10.4 \\
25 & SDF J132347.7+272360 &  4.5 & $<$6.4 &  4.1 & $<$9.0 \\
26 & SDF J132348.9+271530 &  5.0 & $<$6.4 &  4.5 & $<$9.0 \\
27 & SDF J132349.2+273211 &  3.6 & $<$6.4 &  3.3 & $<$9.0 \\
28 & SDF J132349.2+274546 &  6.9 & $<$6.4 &  6.3 & $<$9.0 \\
29 & SDF J132353.4+272602 &  6.7 &   11.3 &  6.1 &   15.9 \\
30 & SDF J132357.1+272448 &  3.3 &    8.0 &  3.0 &   11.2 \\
31 & SDF J132401.5+273837 &  3.6 & $<$6.4 &  3.3 & $<$9.0 \\
32 & SDF J132402.6+274653 &  3.2 &   13.9 &  2.9 &   19.5 \\
33 & SDF J132410.5+272811 &  4.0 & $<$6.4 &  3.7 & $<$9.0 \\
34 & SDF J132419.3+274125 &  5.8 & $<$6.4 &  5.3 & $<$9.0 \\
35 & SDF J132422.6+274459 &  5.2 &    9.2 &  4.7 &   12.9 \\
36 & SDF J132424.2+272649 &  2.6 &    8.8 &  2.4 &   12.3 \\
37 & SDF J132425.4+272410 &  3.2 & $<$6.4 &  2.9 & $<$9.0 \\
38 & SDF J132425.9+274324 &  4.4 & $<$6.4 &  4.0 & $<$9.0 \\
39 & SDF J132425.0+273606 &  4.7 & $<$6.4 &  4.3 & $<$9.0 \\
40 & SDF J132434.3+274056 &  5.1 & $<$6.4 &  4.6 & $<$9.0 \\
41 & SDF J132435.0+273957 &  5.4 & $<$6.4 &  5.0 & $<$9.0 \\
42 & SDF J132436.6+272223 &  3.7 & $<$6.4 &  3.3 & $<$9.0 \\
43 & SDF J132440.2+272553 &  4.6 & $<$6.4 &  4.1 & $<$9.0 \\
44 & SDF J132443.4+272633 &  4.2 &   10.4 &  3.9 &   14.5 \\
45 & SDF J132444.4+273942 &  4.8 & $<$6.4 &  4.4 & $<$9.0 \\
46 & SDF J132445.6+273033 &  5.6 &    8.1 &  5.1 &   11.3 \\
47 & SDF J132447.7+271106 &  4.3 &    9.9 &  3.9 &   13.8 \\
48 & SDF J132449.5+274237 &  2.7 & $<$6.4 &  2.5 & $<$9.0 \\
49 & SDF J132450.7+272160 &  4.6 & $<$6.4 &  4.2 & $<$9.0 \\
50 & SDF J132455.4+271314 &  4.0 & $<$6.4 &  3.7 & $<$9.0 \\
51 & SDF J132455.8+274015 &  3.2 & $<$6.4 &  2.9 & $<$9.0 \\
52 & SDF J132458.5+273913 & 10.4 & $<$6.4 &  9.5 & $<$9.0 \\
53 & SDF J132458.0+272349 &  4.4 & $<$6.4 &  4.0 & $<$9.0 \\
54 & SDF J132503.4+273838 &  3.3 & $<$6.4 &  3.0 & $<$9.0 \\
55 & SDF J132506.4+274047 &  3.1 & $<$6.4 &  2.8 & $<$9.0 \\
56 & SDF J132516.7+272236 &  5.5 & $<$6.4 &  5.0 & $<$9.0 \\
57 & SDF J132528.0+271328 &  3.9 & $<$6.4 &  3.5 & $<$9.0 \\
58 & SDF J132533.4+271420 &  5.4 & $<$6.4 &  4.9 & $<$9.0 \\
\hline
\multicolumn{6}{l}{
\footnotemark[*]{The UV luminosity at $\lambda$ = 1260 \AA.}
}\\
\end{longtable}
}



\clearpage    

\begin{figure}
  \begin{center}
    \FigureFile(80mm,80mm){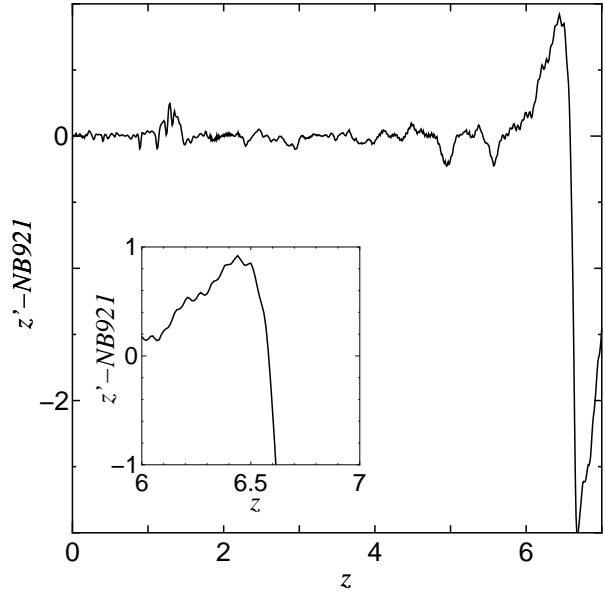}
  \end{center}
\caption{
The $z^\prime - NB921$ color for a galaxy without a Ly$\alpha$ emission line
shown as a function of the redshift. The inset shows a close up
for a redshift range between $z=6.0$ and $z=7.0$.
\label{fig:fig1}
}
\end{figure}

\clearpage

\begin{figure}
  \begin{center}
    \FigureFile(80mm,80mm){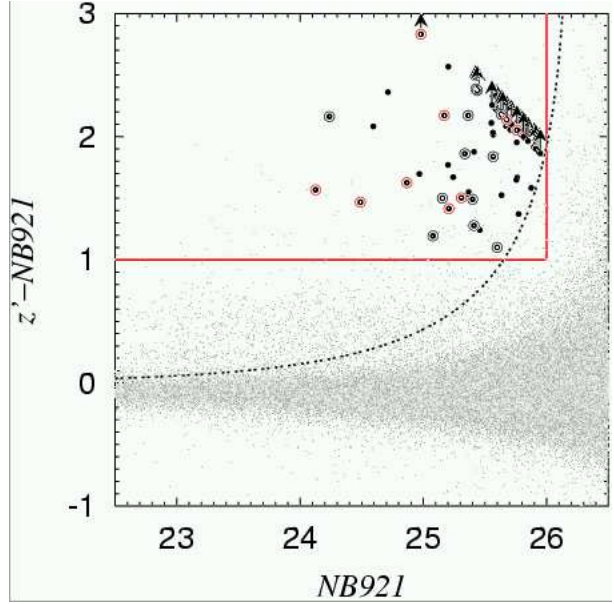}
  \end{center}
\caption{
Color--magnitude diagram between $z^{\prime} - NB921$ and $NB921$.
All objects detected down to the apparent magnitude of $NB921=26.5$
in the $NB921$-selected catalog are shown.
The horizontal red solid line corresponds to the color of
$z^{\prime} - NB921=1.0$ and the vertical red one corresponds to
the 5$\sigma$ limiting magnitude of $NB921=26.0$.
The dashed curve shows the distribution of $3 \sigma$ error in the color of
$z^{\prime} - NB921$.
The 58 LAE candidates in the photometric sample are shown by filled circles,
the twenty ones in the spectroscopic sample by filled circles with a large open circle,
while the nine LAEs confirmed by our spectroscopy (see subsection 3.1)
are shown with the same symbol but with a red large open circle.
Data points without detection at $z^\prime$ are shown with a upper arrow;
note that one $\sigma$ $z^\prime$ magnitude is adopted.
}\label{fig:fig2}
\end{figure}

\clearpage

\begin{figure}
  \begin{center}
    \FigureFile(80mm,80mm){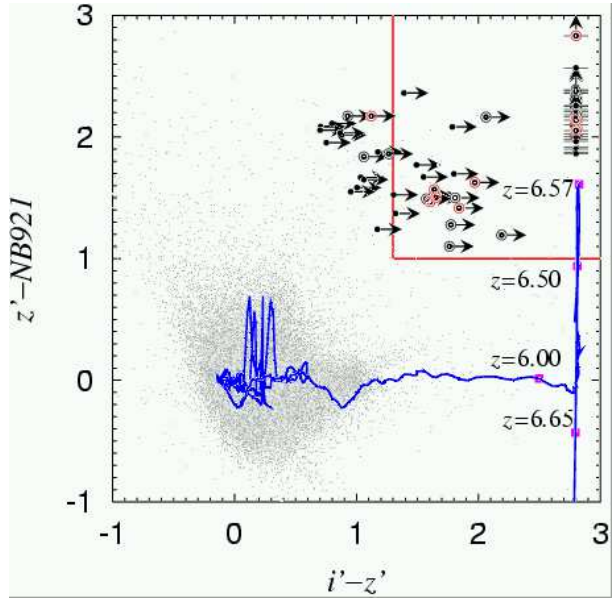}
  \end{center}
\caption{
Diagram between $z^{\prime} - NB921$ and
$i^{\prime} - z^{\prime}$. The meanings of the symbols are the same
as those in figure 2. See the text for the color evolution of a model galaxy
(blue curve).
}\label{fig:fig3}
\end{figure}

\clearpage

\begin{figure}
  \begin{center}
    \FigureFile(80mm,80mm){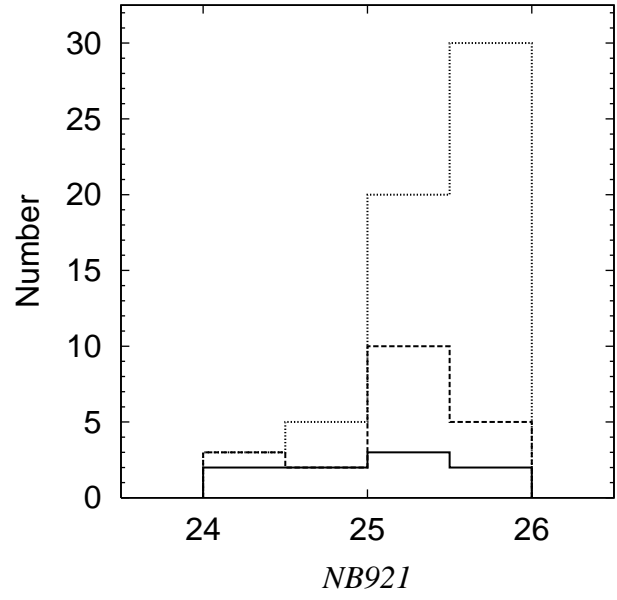}
  \end{center}
\caption{
Distributions of $NB921$ magnitudes for
the photometric sample (58 objects; dotted histograms), 
the spectroscopic sample (20 objects; dashed histograms),
and the spectroscopically-confirmed LAE sample (9 objects;
solid histograms).
}\label{fig:fig4}
\end{figure}

\clearpage

\begin{figure}
  \begin{center}
    \FigureFile(80mm,80mm){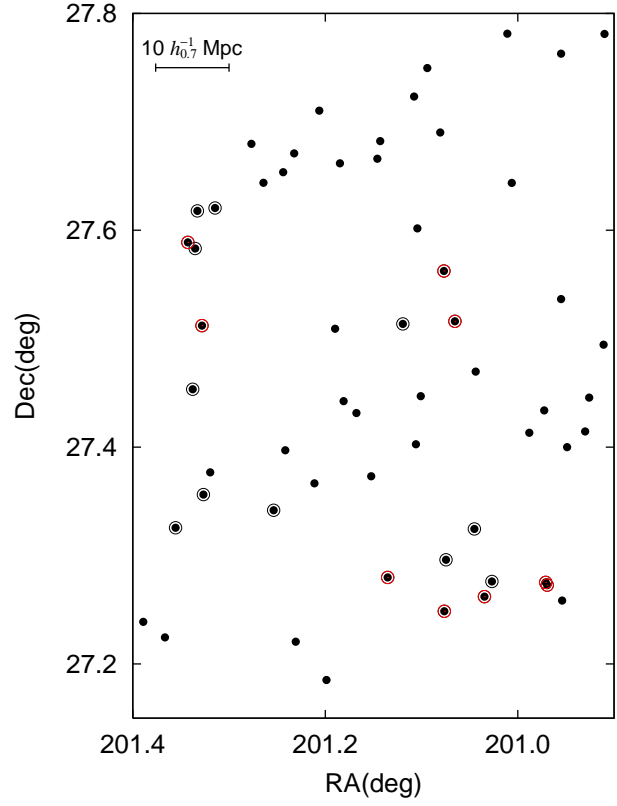}
  \end{center}
\caption{
Spatial distribution of the 58 objects given in our photometric sample
(see table 2). The 20 LAE candidates in the spectroscopic sample are
shown with a black open circle. The nine confirmed LAEs are shown
with a red large open circle.
\label{fig:fig5}
}
\end{figure}

\clearpage

\begin{figure}
  \begin{center}
    \FigureFile(80mm,80mm){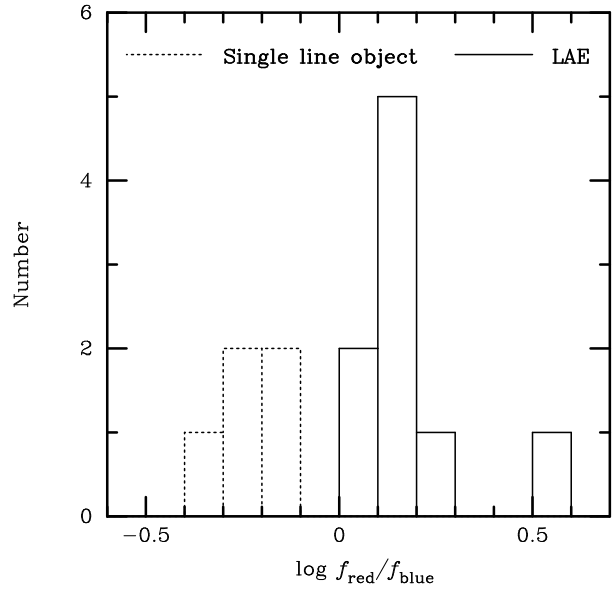}
  \end{center}
\caption{
Frequency distribution of log $f_{\rm red}$/$f_{\rm blue}$ for
 the sample of 9 LAEs (solid-line) and 5 single-line objects
(dotted-line).
}\label{fig:fig6}
\end{figure}

\clearpage

\setcounter{figure}{0}

\begin{figure}
\renewcommand{\figurename}{Fig. 7-}
  \begin{center}
    \FigureFile(80mm,80mm){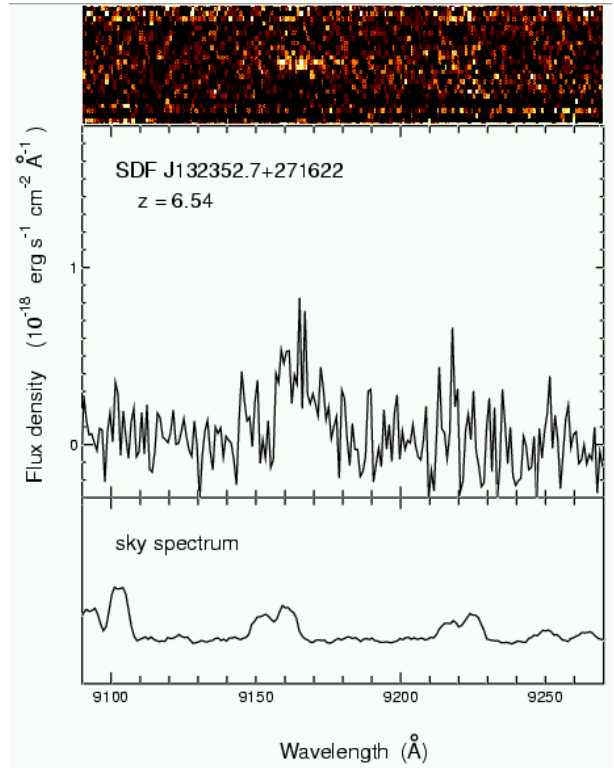}
  \end{center}
\caption{
Observed optical spectra of nine LAE candidates
(No.~1~SDF J132352.7+271622).
}\label{fig:fig7-1}
\end{figure}

\clearpage

\begin{figure}
\renewcommand{\figurename}{Fig. 7-}
  \begin{center}
    \FigureFile(80mm,80mm){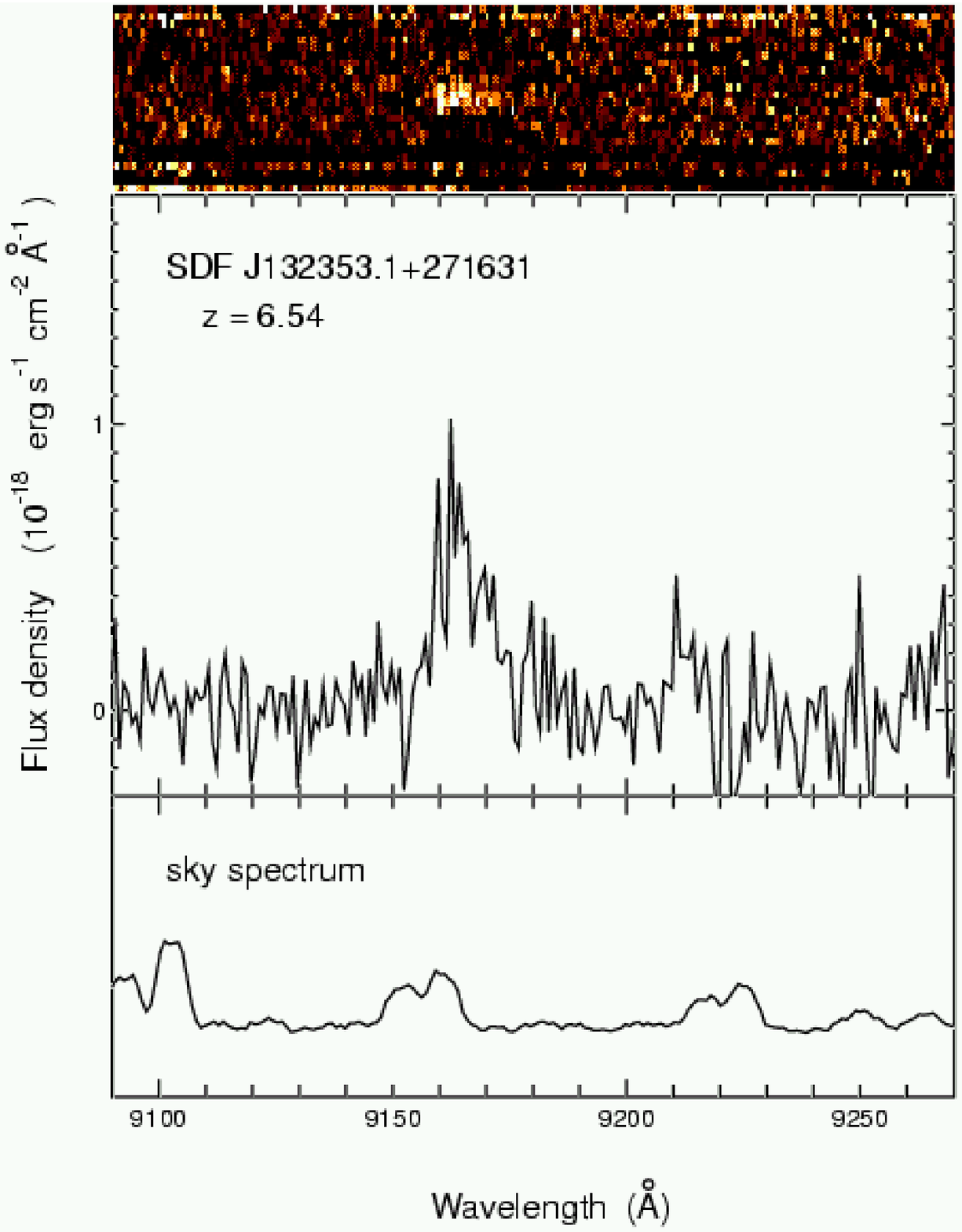}
  \end{center}
\caption{
Observed optical spectra of nine LAE candidates
(No.~2~SDF J132353.1+271631).
}\label{fig:fig7-2}
\end{figure}

\clearpage

\begin{figure}
\renewcommand{\figurename}{Fig. 7-}
  \begin{center}
    \FigureFile(80mm,80mm){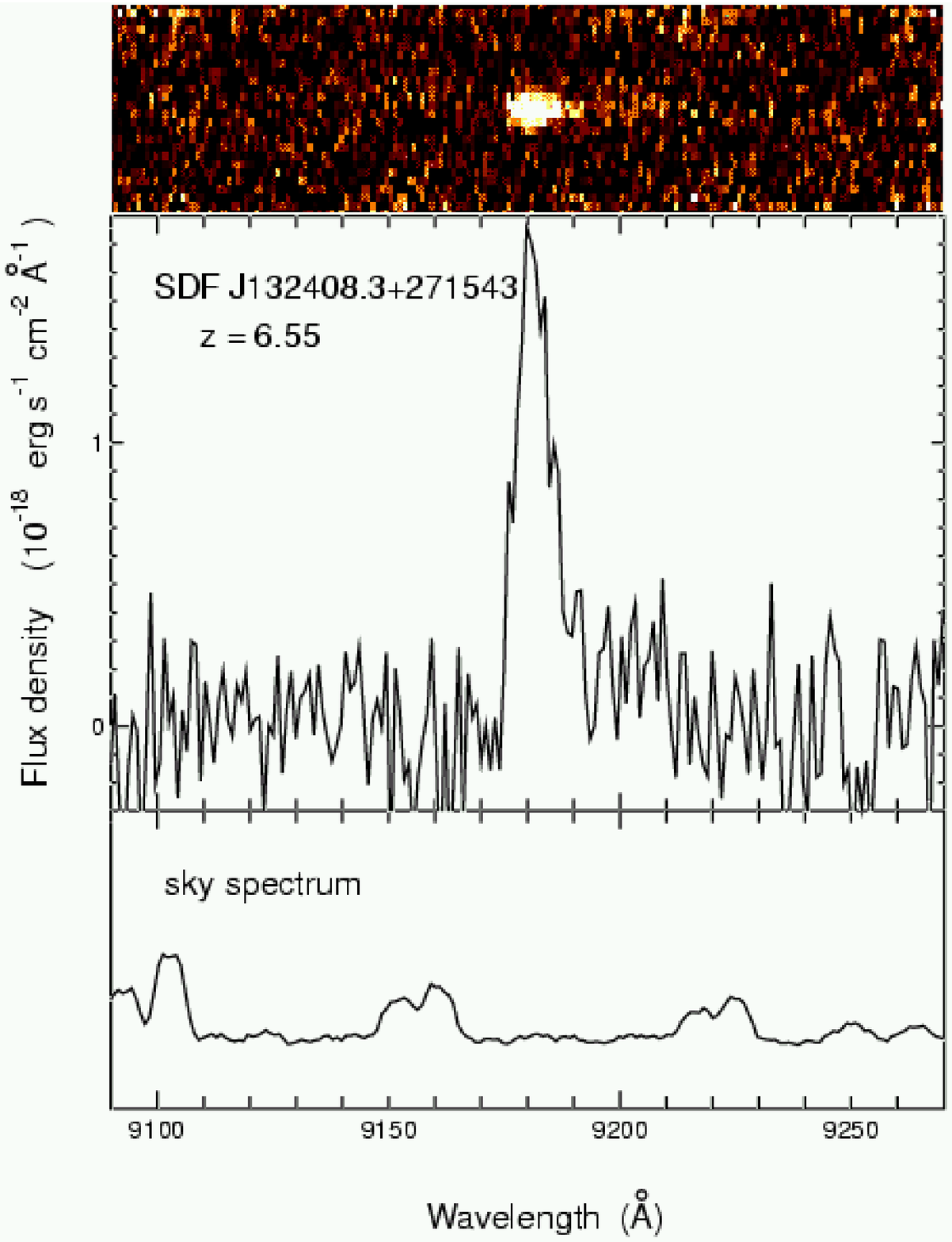}
  \end{center}
\caption{
Observed optical spectra of nine LAE candidates
(No.~3~SDF J132408.3+271543).
}\label{fig:fig7-3}
\end{figure}

\clearpage

\begin{figure}
\renewcommand{\figurename}{Fig. 7-}
  \begin{center}
    \FigureFile(80mm,80mm){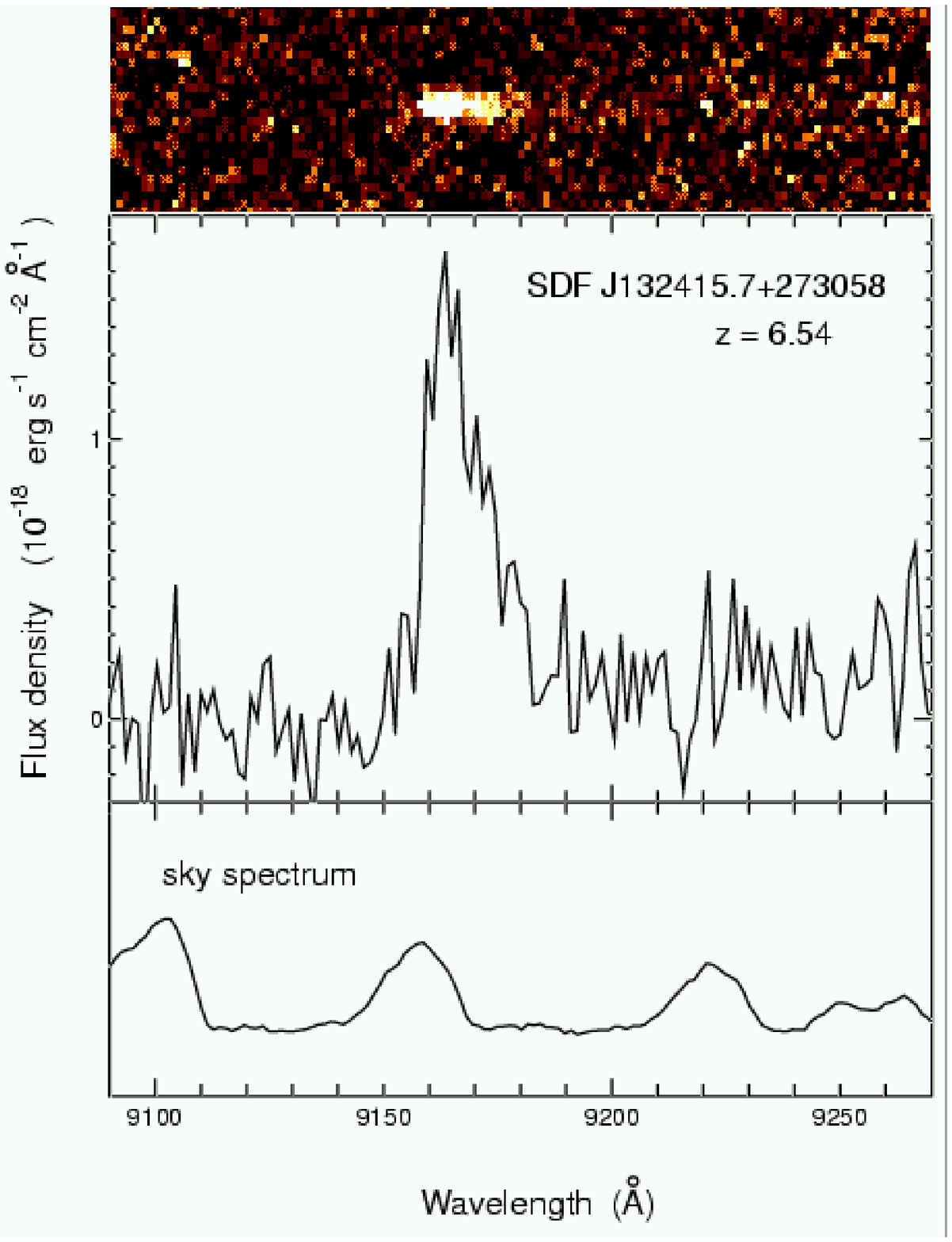}
  \end{center}
\caption{
Observed optical spectra of nine LAE candidates
(No.~4~SDF J132415.7+273058). This object was already reported in
Kodaira et al. (2003).
}\label{fig:fig7-4}
\end{figure}

\clearpage

\begin{figure}
\renewcommand{\figurename}{Fig. 7-}
  \begin{center}
    \FigureFile(80mm,80mm){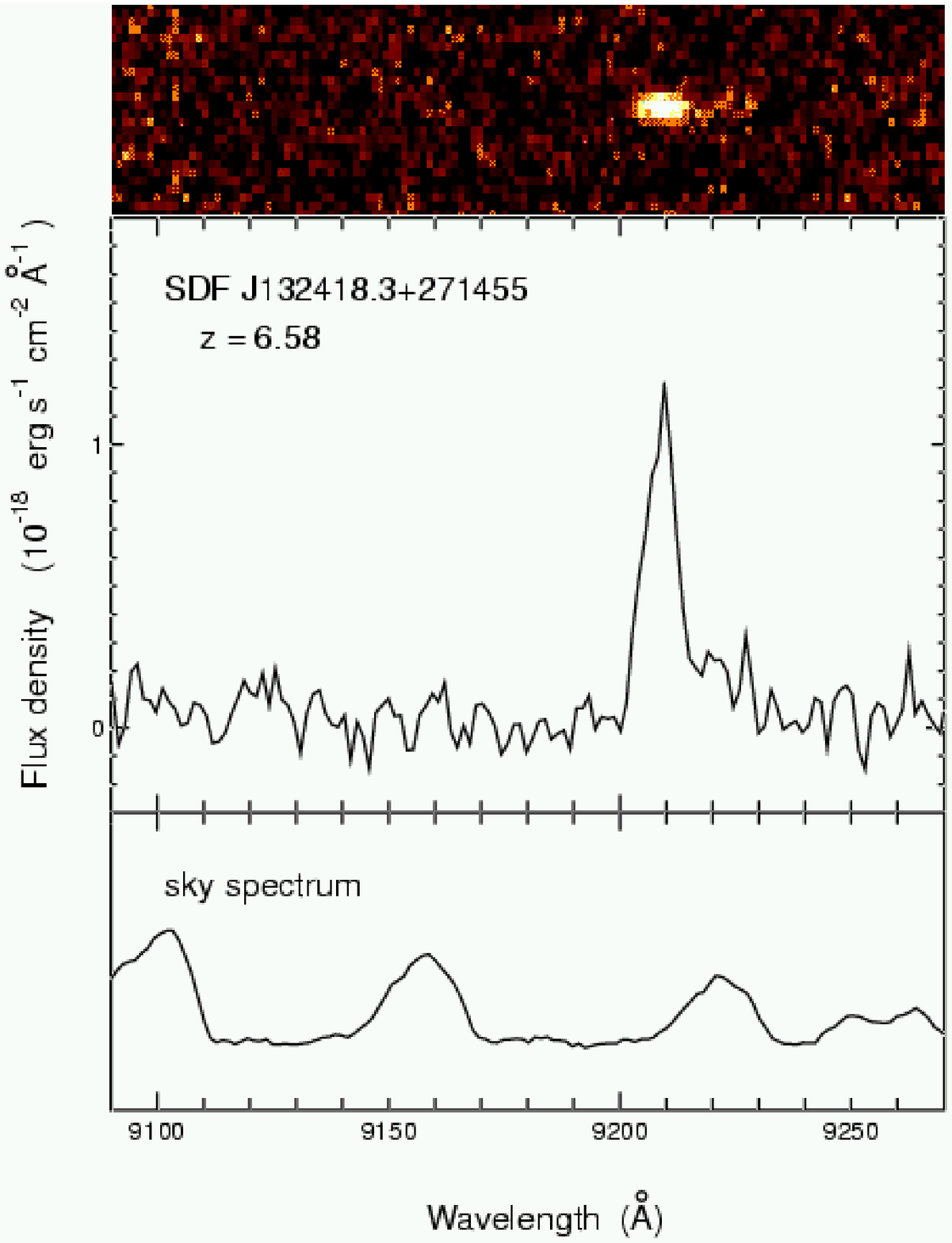}
  \end{center}
\caption{
Observed optical spectra of nine LAE candidates
(No.~5~SDF J132418.3+271455). This object was already reported in
Kodaira et al. (2003).
}\label{fig:fig7-5}
\end{figure}

\clearpage

\begin{figure}
\renewcommand{\figurename}{Fig. 7-}
  \begin{center}
    \FigureFile(80mm,80mm){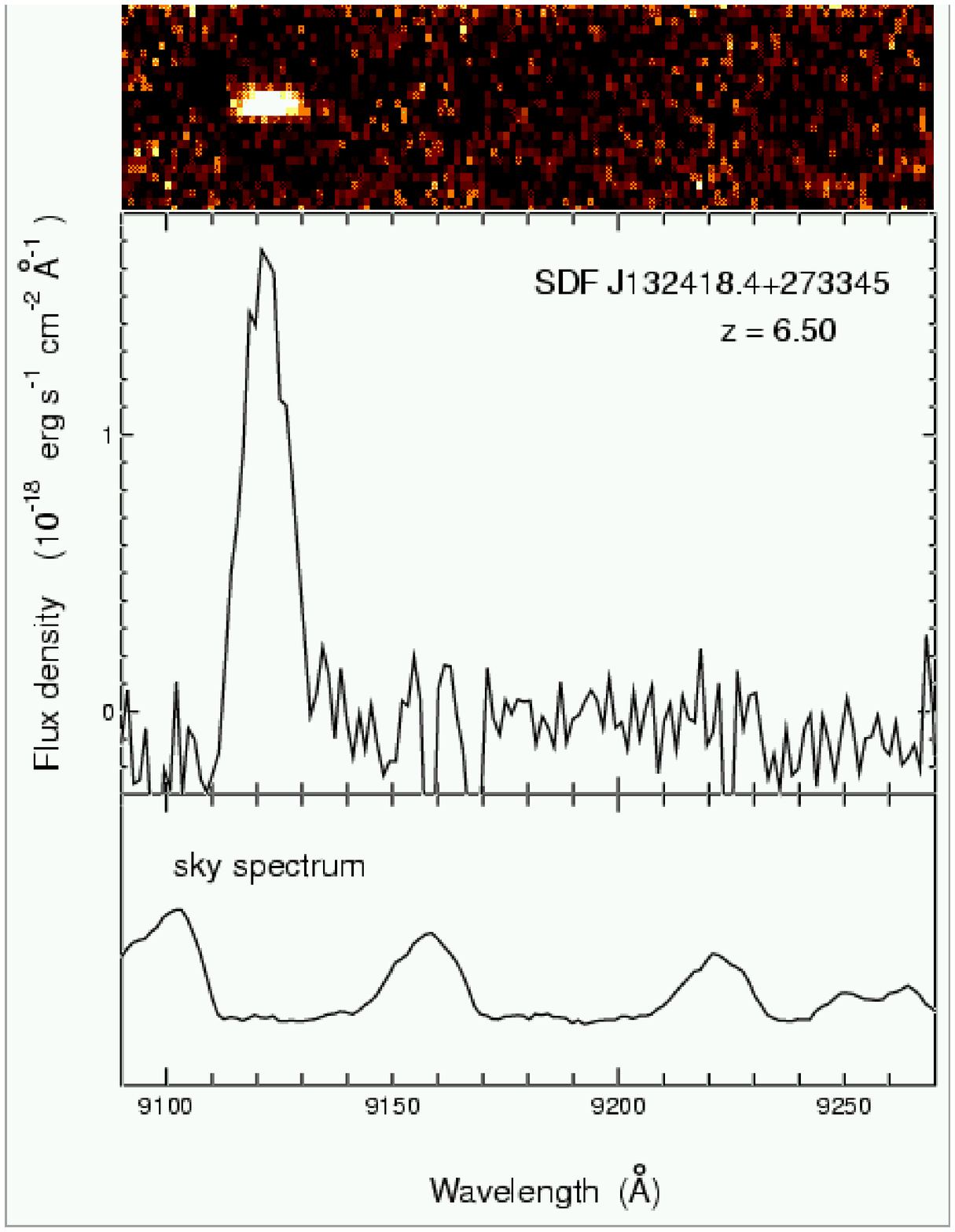}
  \end{center}
\caption{
Observed optical spectra of nine LAE candidates
(No.~6~SDF J132418.4+273345).
}\label{fig:fig7-6}
\end{figure}

\clearpage

\begin{figure}
\renewcommand{\figurename}{Fig. 7-}
  \begin{center}
    \FigureFile(80mm,80mm){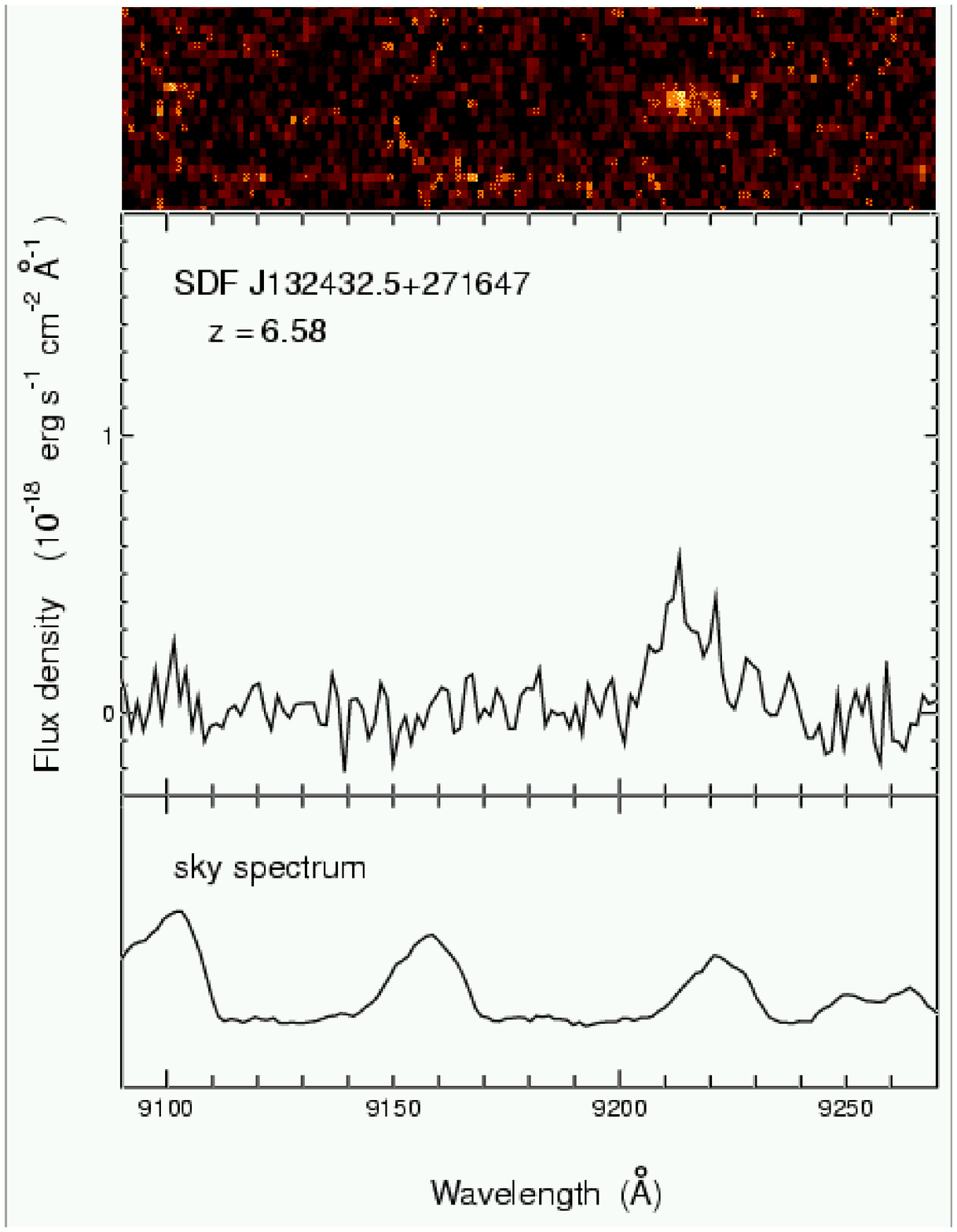}
  \end{center}
\caption{
Observed optical spectra of nine LAE candidates
(No.~7~SDF J132432.5+271647).
}\label{fig:fig7-7}
\end{figure}

\clearpage

\begin{figure}
\renewcommand{\figurename}{Fig. 7-}
  \begin{center}
    \FigureFile(80mm,80mm){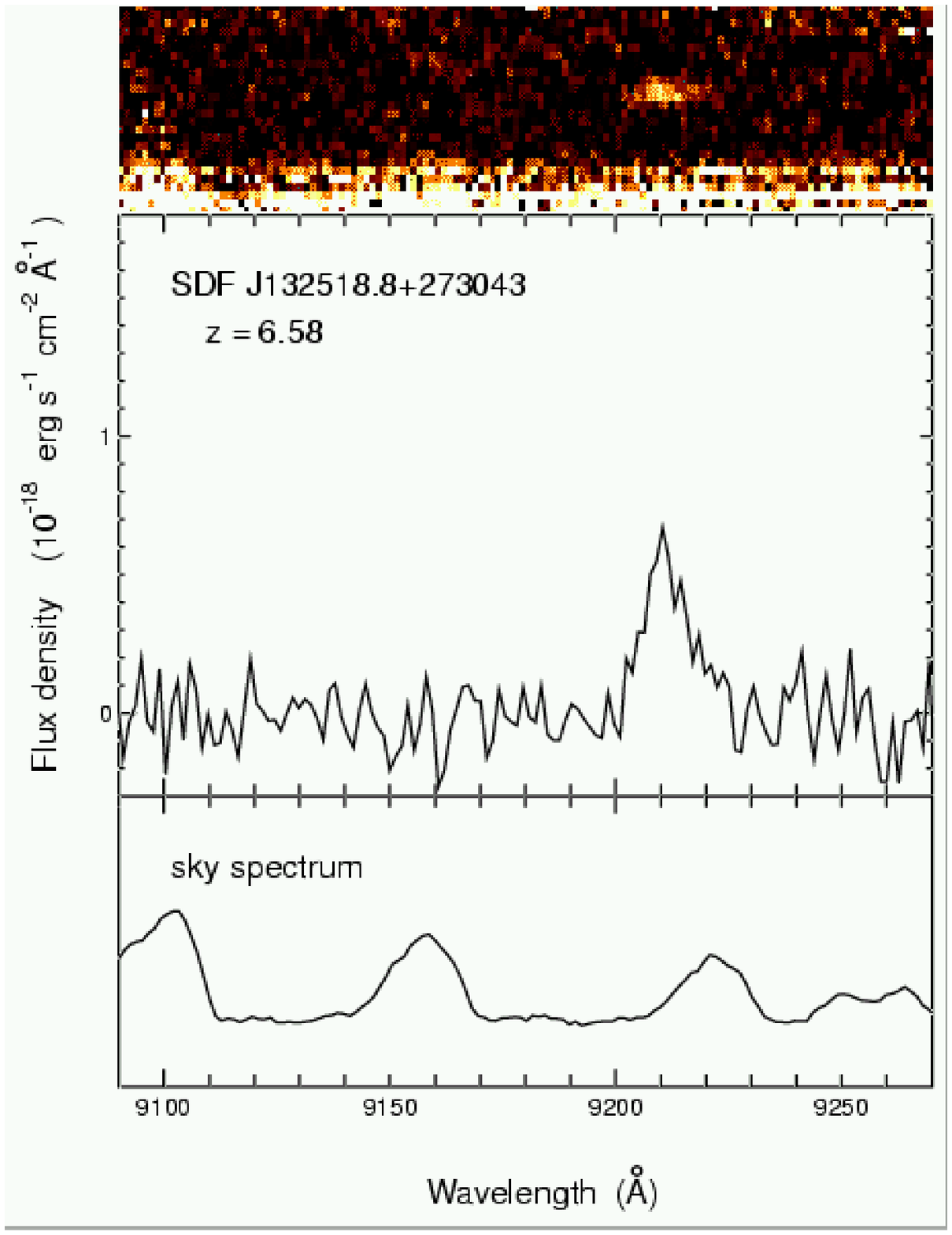}
  \end{center}
\caption{
Observed optical spectra of nine LAE candidates
(No.~8~SDF J132518.8+273043).
}\label{fig:fig7-8}
\end{figure}

\clearpage

\begin{figure}
\renewcommand{\figurename}{Fig. 7-}
  \begin{center}
    \FigureFile(80mm,80mm){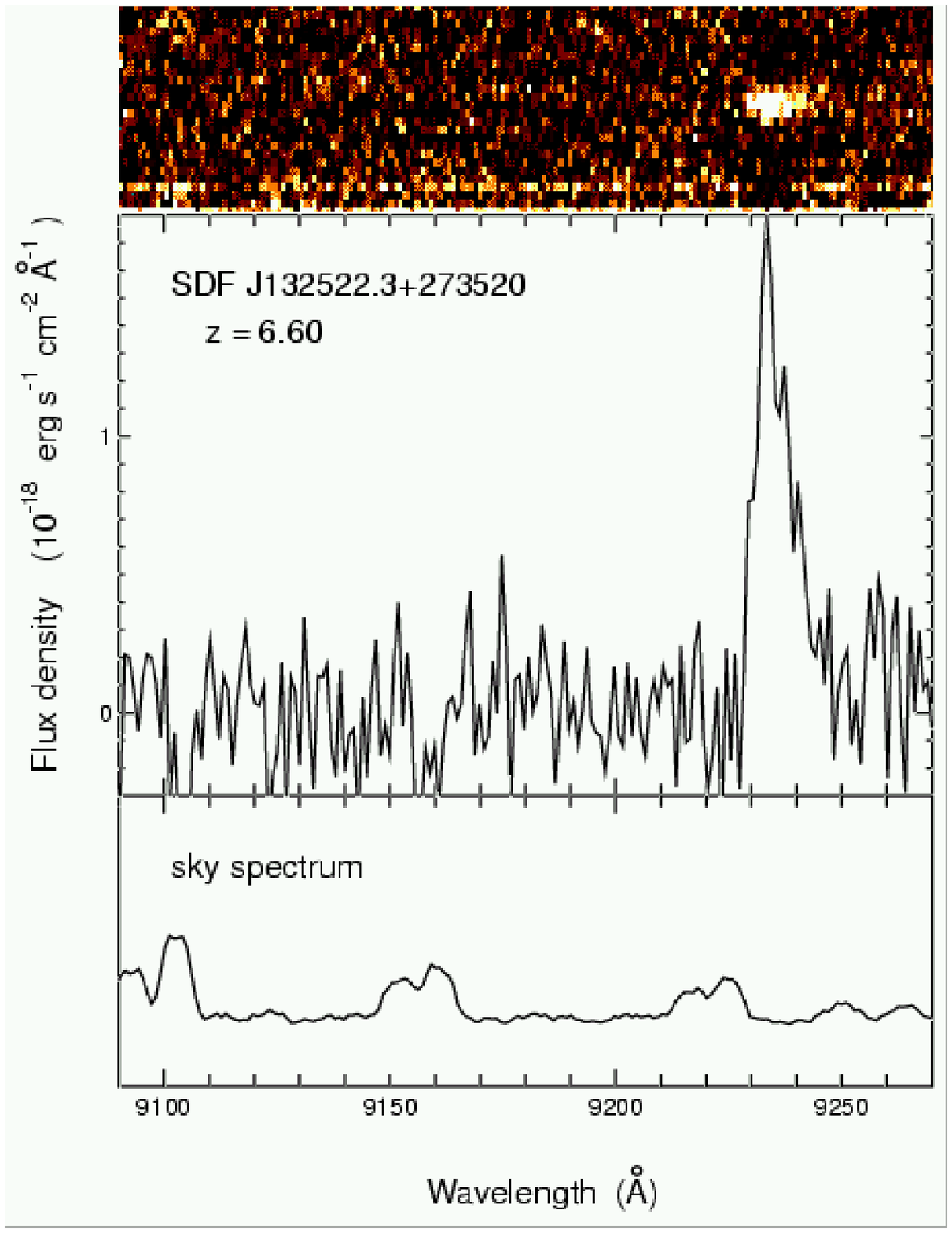}
  \end{center}
\caption{
Observed optical spectra of nine LAE candidates
(No.~9~SDF J132522.3+273520).
}\label{fig:fig7-9}
\end{figure}

\clearpage

\setcounter{figure}{0}

\begin{figure}
\renewcommand{\figurename}{Fig. 8-}
  \begin{center}
    \FigureFile(80mm,80mm){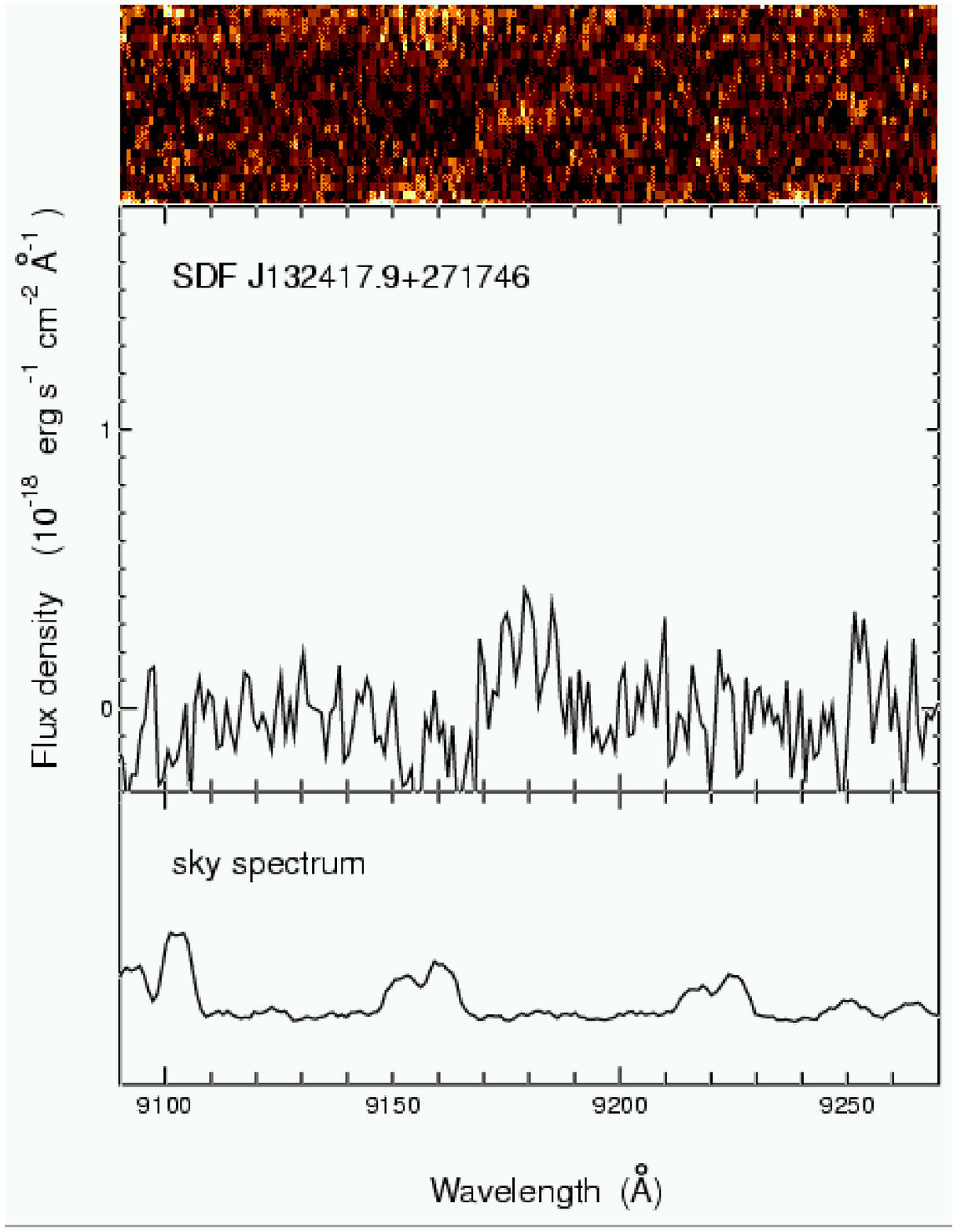}
  \end{center}
\caption{
Observed optical spectra of five single-line objects
(No.~12~SDF J132417.9+271746).
}\label{fig:fig8-1}
\end{figure}

\clearpage

\begin{figure}
\renewcommand{\figurename}{Fig. 8-}
  \begin{center}
    \FigureFile(80mm,80mm){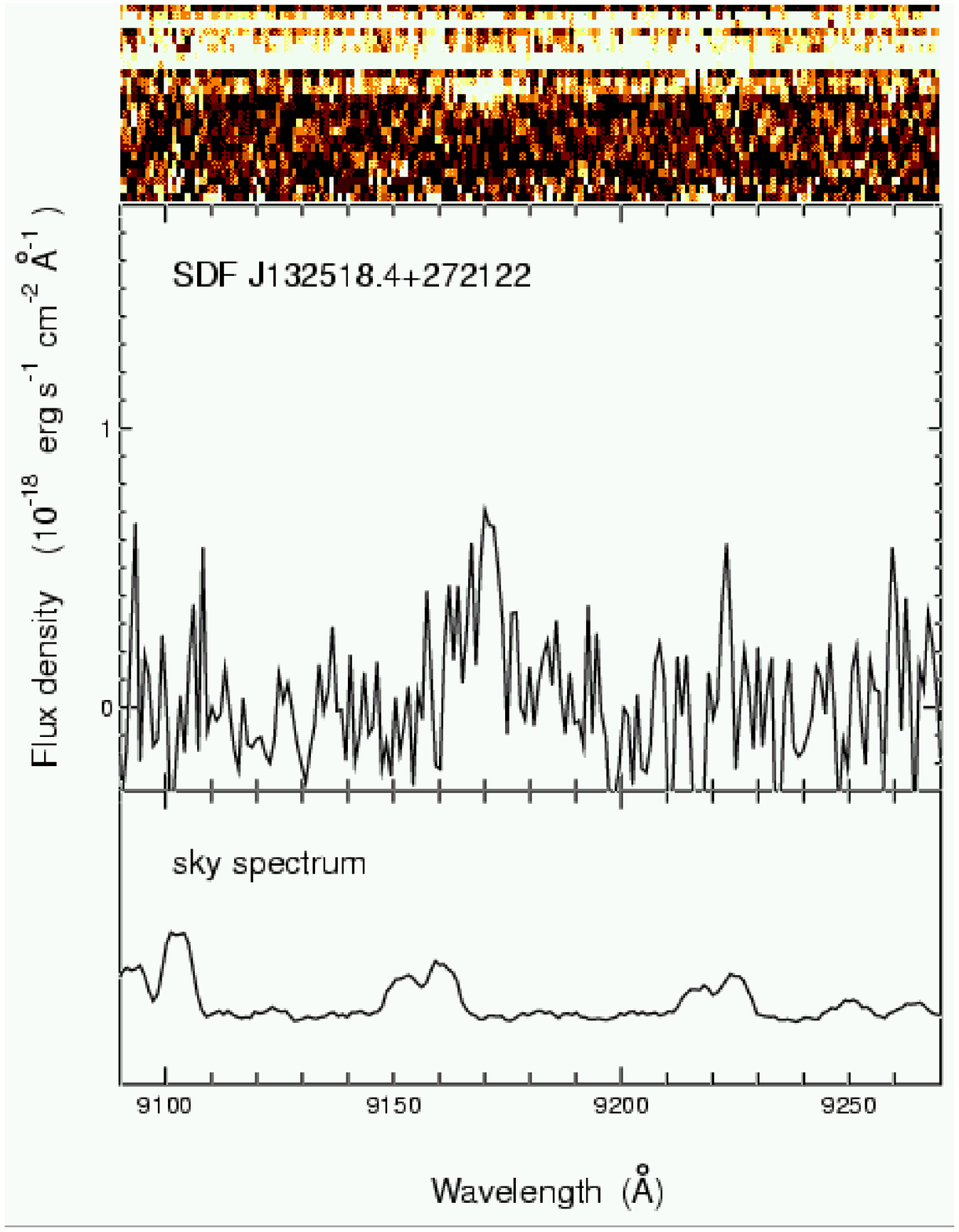}
  \end{center}
\caption{
Observed optical spectra of five single-line objects
(No.~16~SDF J132518.4+272122).
}\label{fig:fig8-2}
\end{figure}

\clearpage

\begin{figure}
\renewcommand{\figurename}{Fig. 8-}
  \begin{center}
    \FigureFile(80mm,80mm){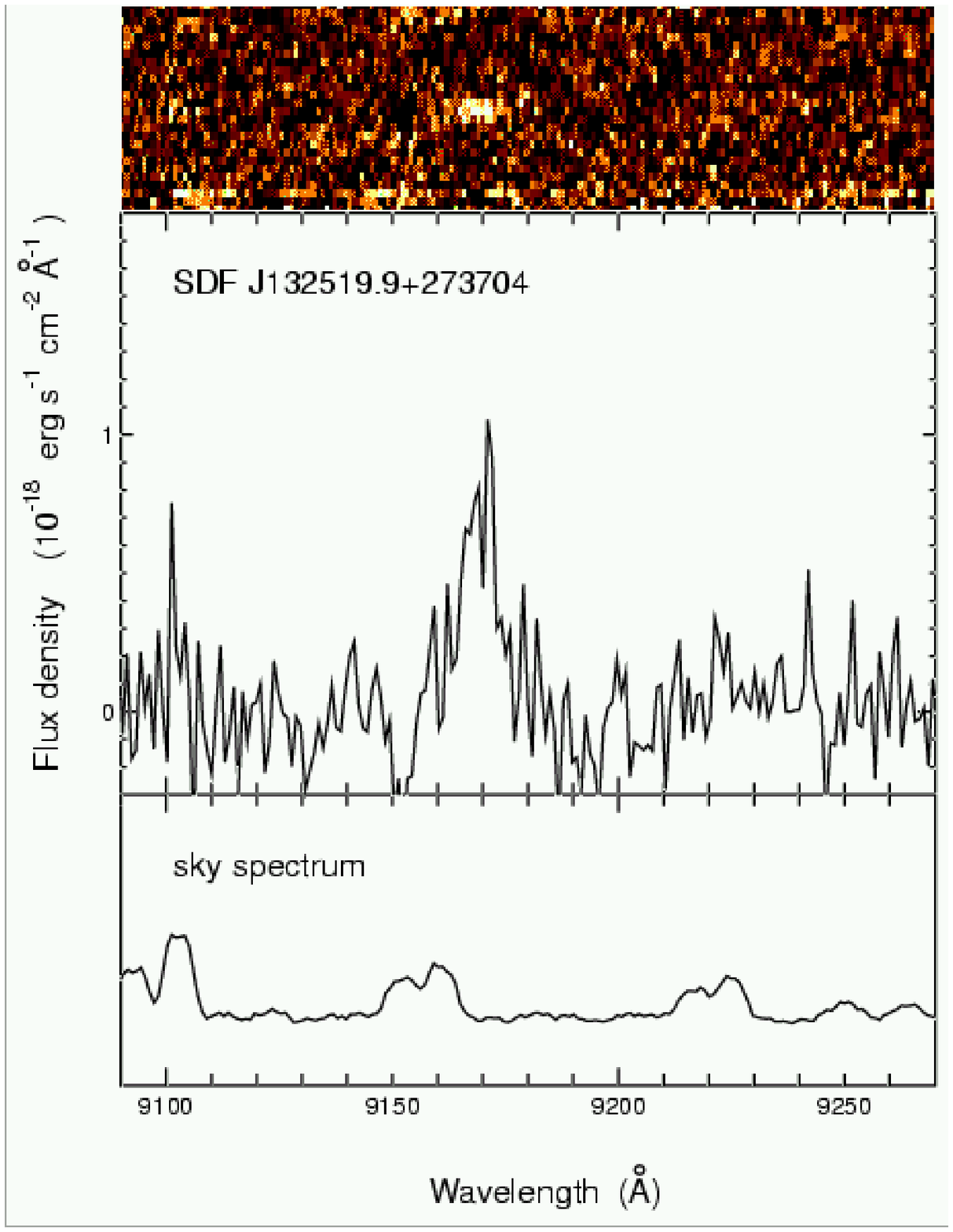}
  \end{center}
\caption{
Observed optical spectra of five single-line objects
(No.~17~SDF J132519.9+273704).
}\label{fig:fig8-3}
\end{figure}

\clearpage

\begin{figure}
\renewcommand{\figurename}{Fig. 8-}
  \begin{center}
    \FigureFile(80mm,80mm){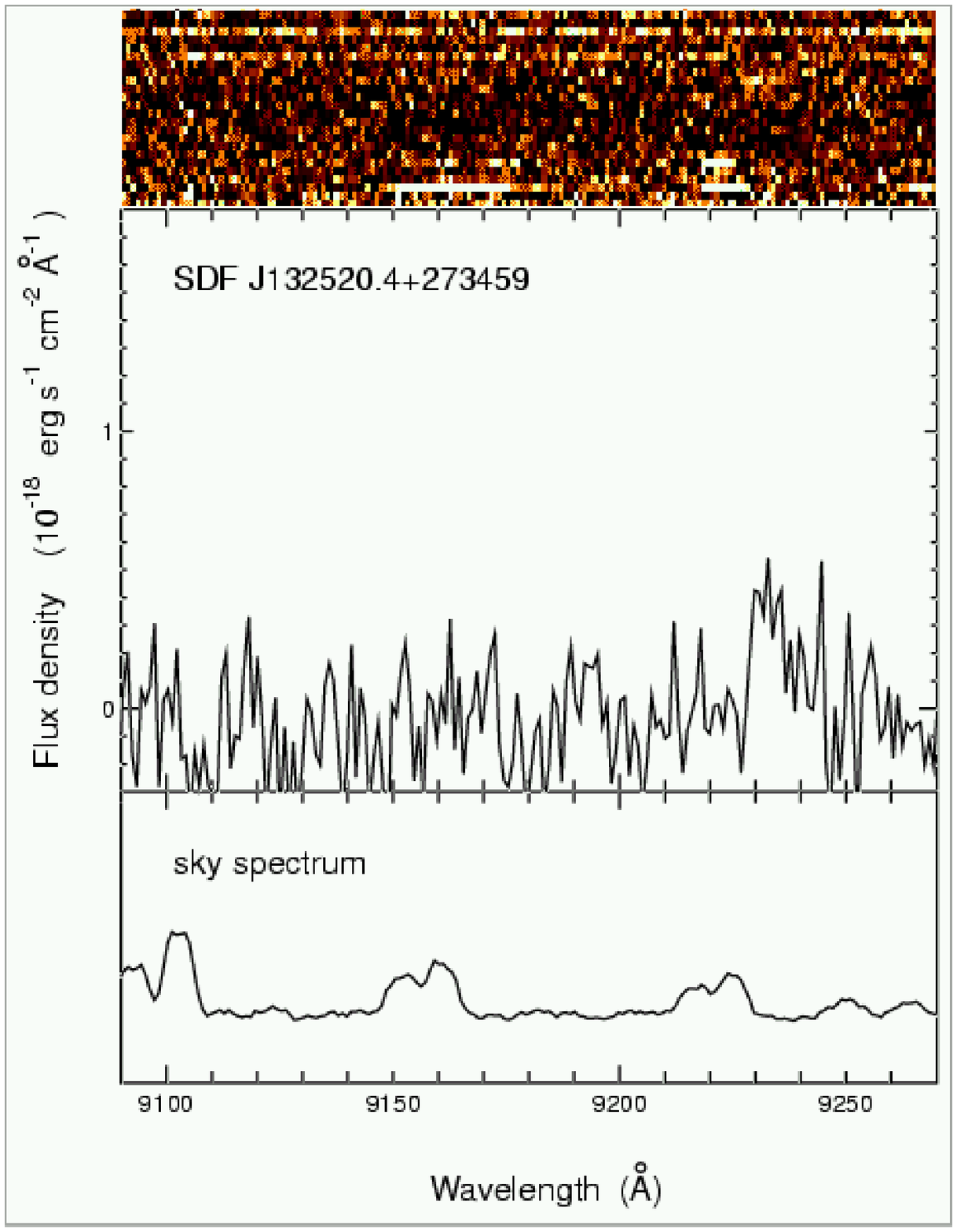}
  \end{center}
\caption{
Observed optical spectra of five single-line objects
(No.~18~SDF J132520.4+273459).
}\label{fig:fig8-4}
\end{figure}

\clearpage

\begin{figure}
\renewcommand{\figurename}{Fig. 8-}
  \begin{center}
    \FigureFile(80mm,80mm){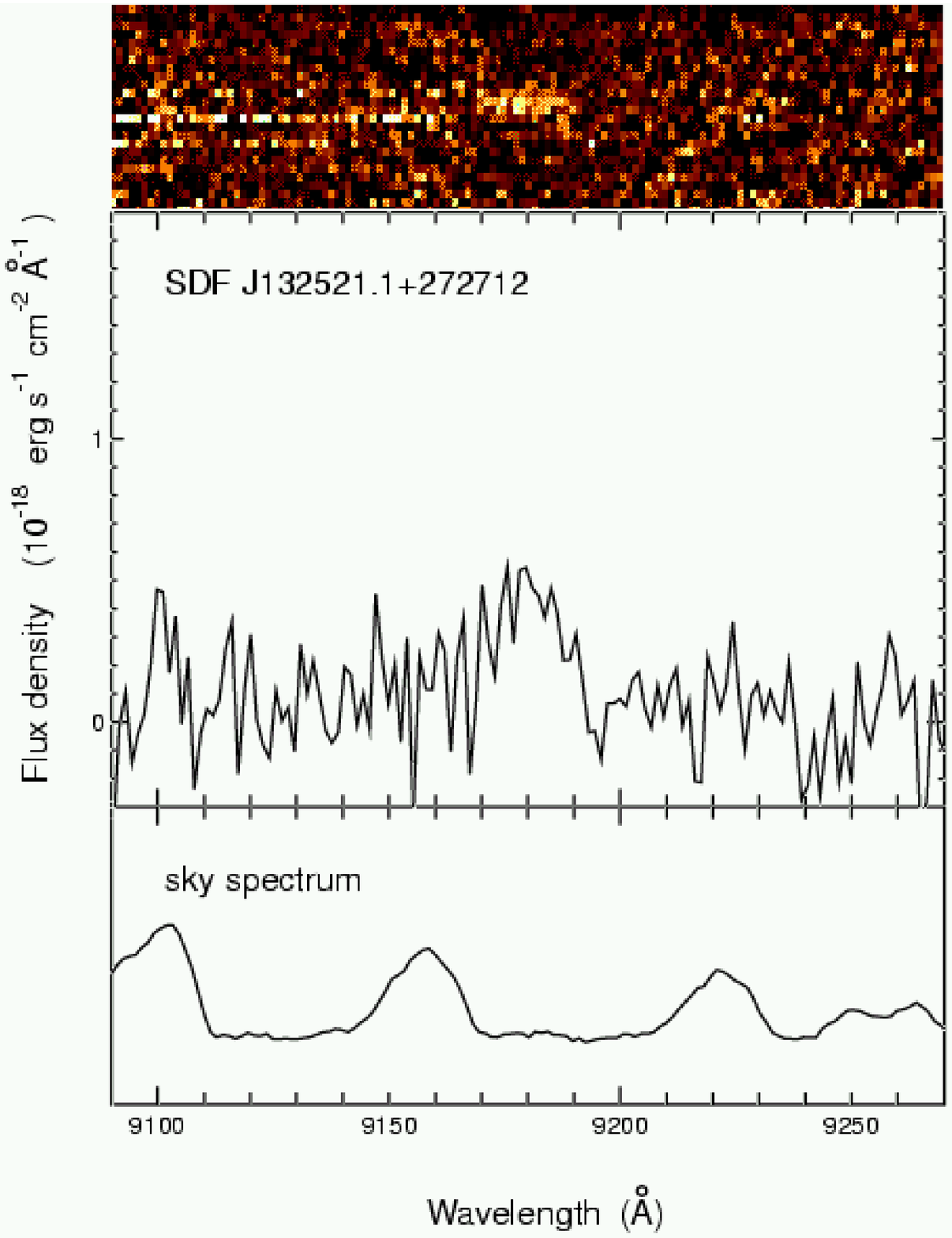}
  \end{center}
\caption{
Observed optical spectra of five single-line objects
(No.~19~SDF J132521.1+272712).
}\label{fig:fig8-5}
\end{figure}

\clearpage

\setcounter{figure}{8}

\begin{figure}
  \begin{center}
    \FigureFile(80mm,80mm){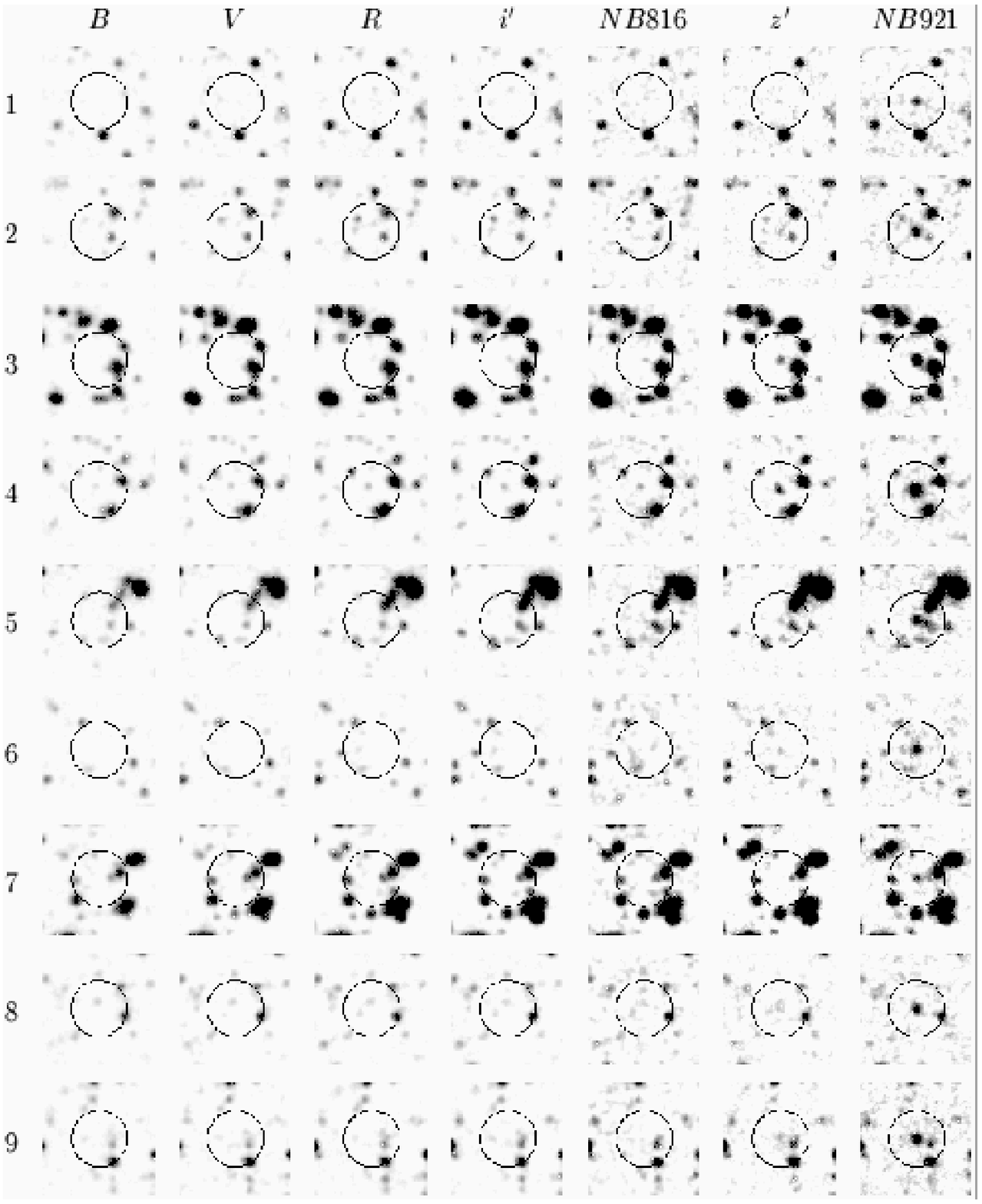}
  \end{center}
\caption{
Thumbnail images of the four LAE candidates.
North is up and east is left. The field of view is 16$''$ $\times$ 16$''$, 
and the diameter of the circle is 8$''$.
}\label{fig:fig9}
\end{figure}

\clearpage

\begin{figure}
  \begin{center}
    \FigureFile(80mm,80mm){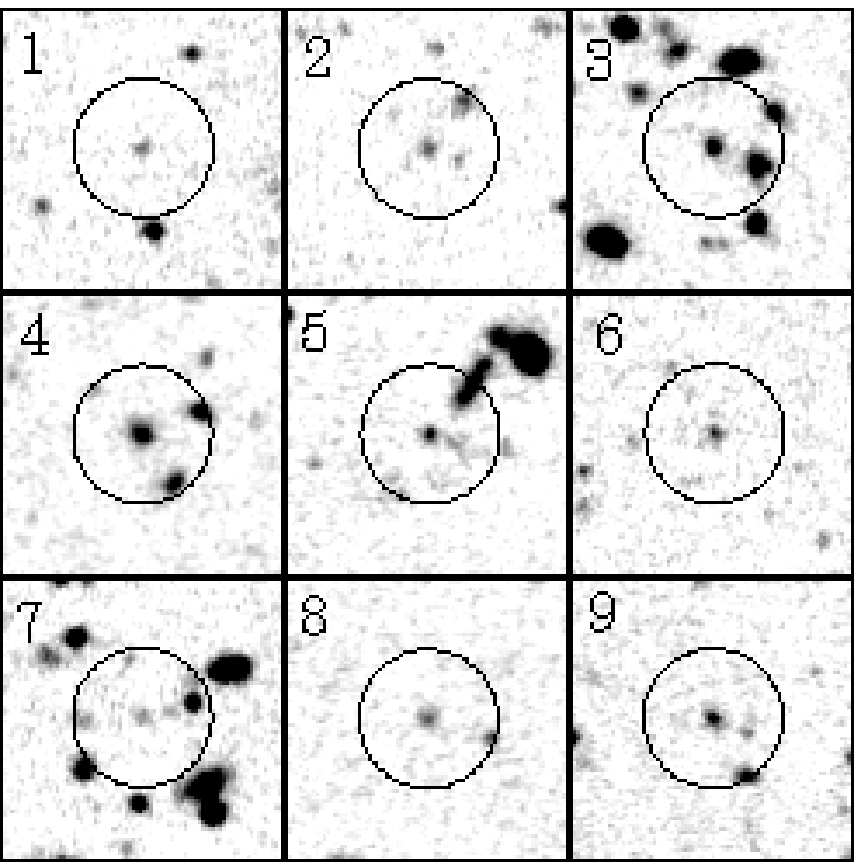}
  \end{center}
\caption{
High-resolution (PSF = 0.$''$71) NB921 images of the nine LAEs.
The number given in each panel corresponds to that given in tables 2, 4, 5, 6, and 7.
}\label{fig:fig10}
\end{figure}

\clearpage

\begin{figure}
  \begin{center}
    \FigureFile(80mm,80mm){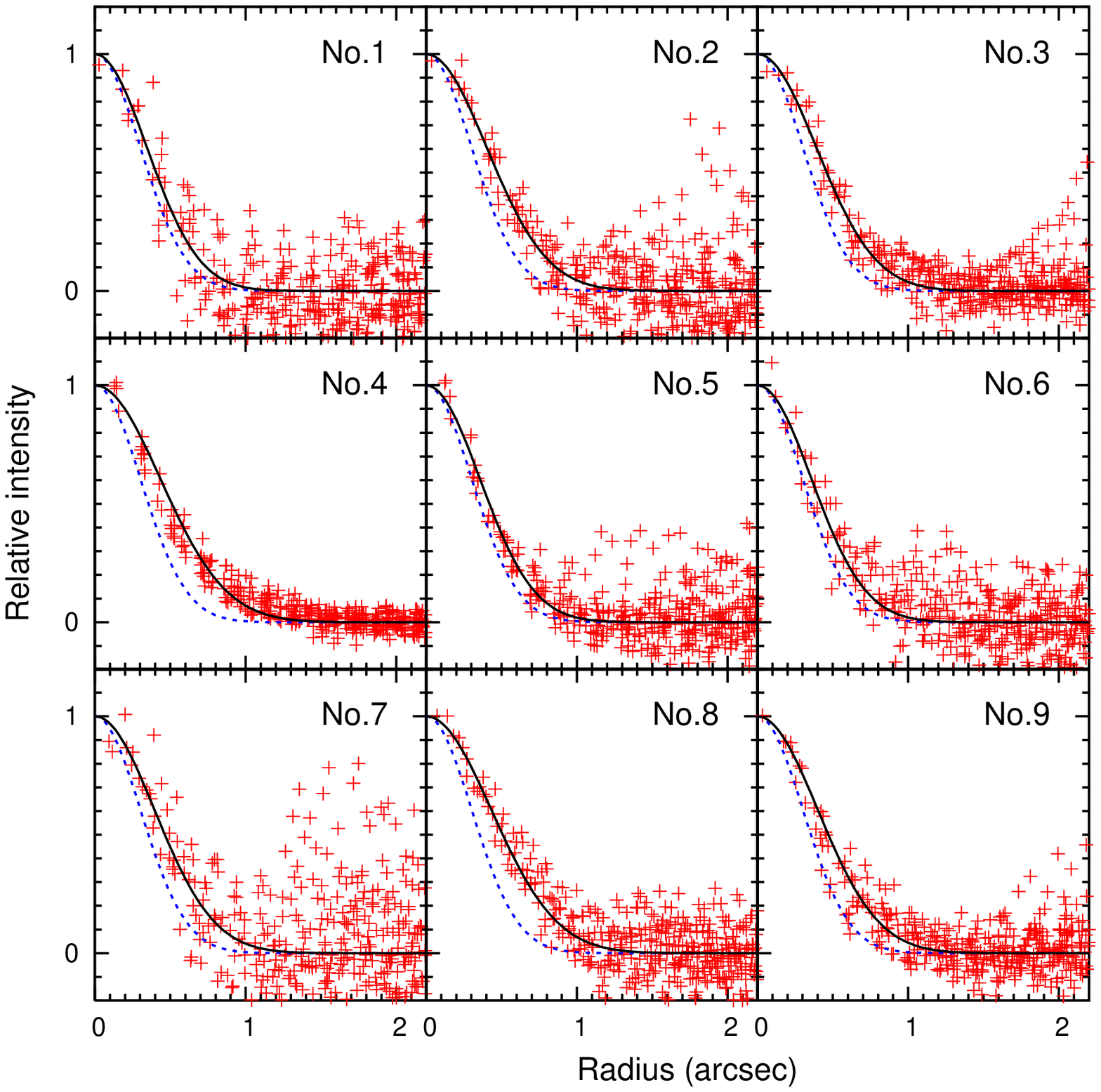}
  \end{center}
\caption{
Surface brightness radial distribution of the nine LAEs.
The point-spread function is shown by a dashed blue curve and 
the best fit profile by black one in each panel.
The number given in each panel corresponds to that given in tables 2, 4, 5, 6, and 7.
}\label{fig:fig11}
\end{figure}

\clearpage

\begin{figure}
  \begin{center}
    \FigureFile(80mm,80mm){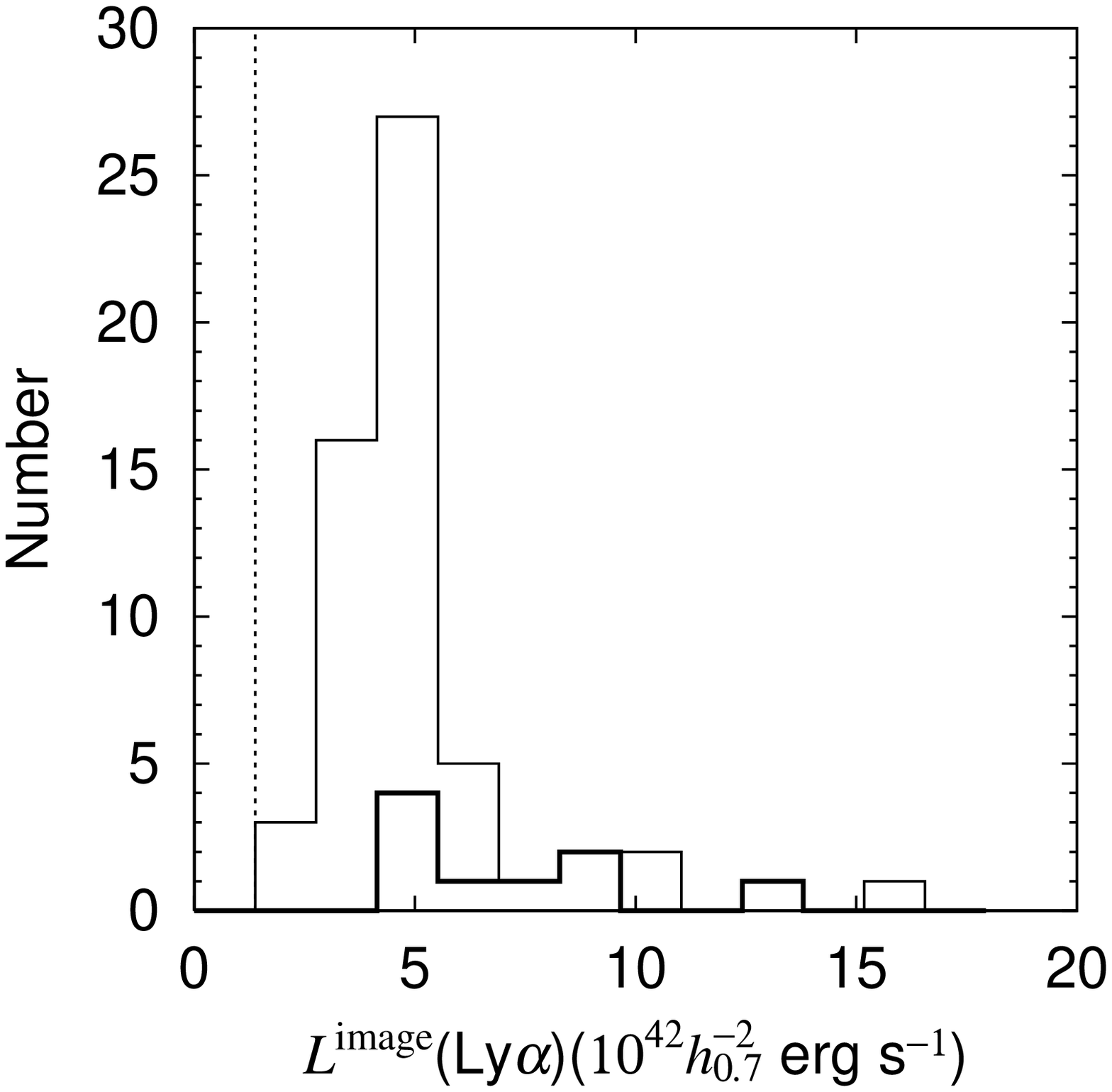}
  \end{center}
\caption{
Histograms of $L^{\rm image}$(Ly$\alpha$) for the photometric sample (58 LAE
candidates) and for the nine LAEs.
The latter is shown by the thick line. The bins correspond to 1$\sigma$
($= 1.4 \times 10^{42} ~ h_{0.7}^{-2}$ erg s$^{-1}$),
2$\sigma$, 3$\sigma$, and 4$\sigma$, of $L^{\rm image}$(Ly$\alpha$);
see table 7 in more details.
}\label{fig:fig12}
\end{figure}

\clearpage

\begin{figure}
  \begin{center}
    \FigureFile(80mm,80mm){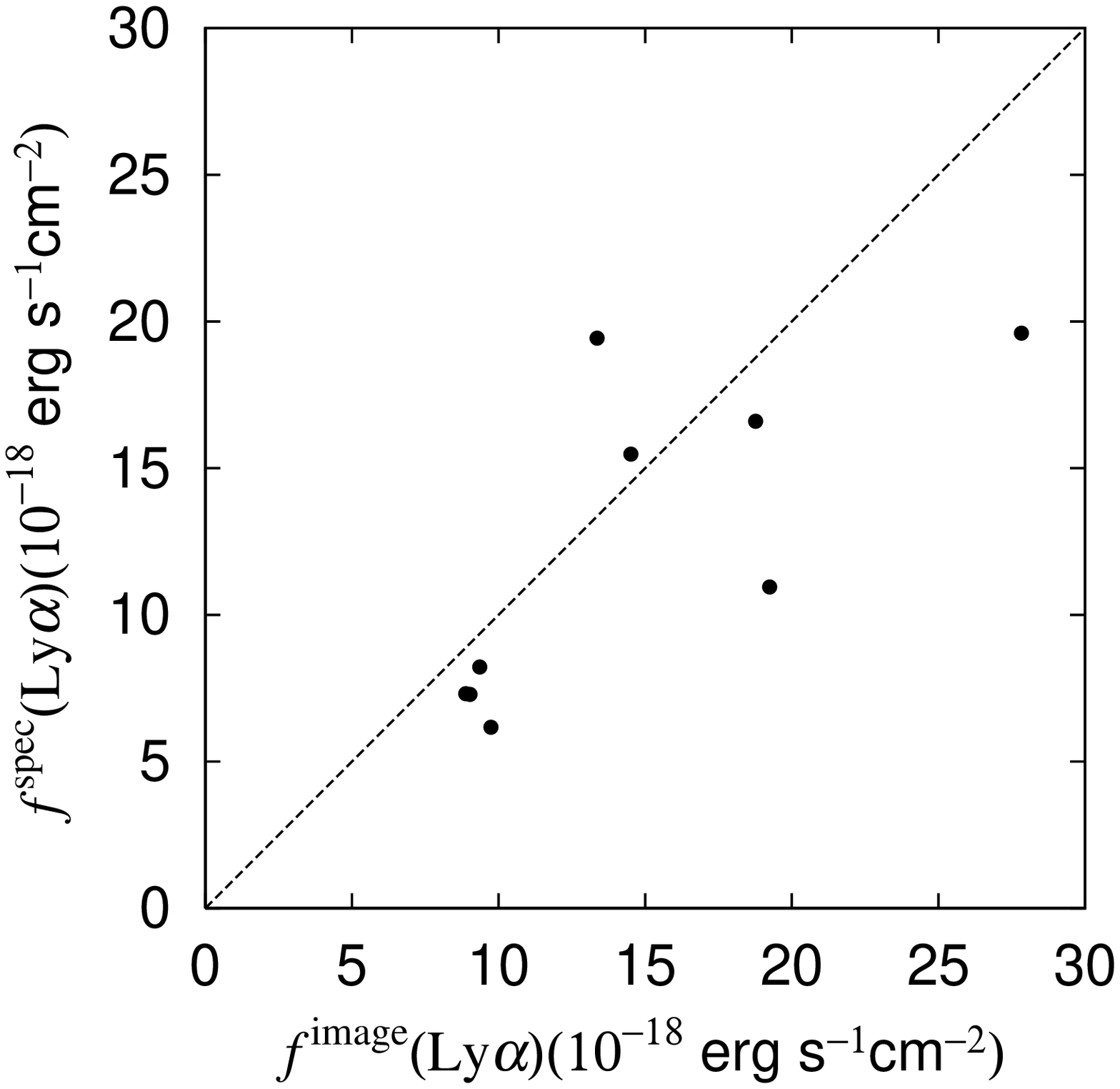}
  \end{center}
\caption{
Comparisons between $f^{\rm spec}$(Ly$\alpha$) and  $f^{\rm image}$(Ly$\alpha$) 
for the nine LAEs.
}\label{fig:fig13}
\end{figure}

\clearpage

\begin{figure}
  \begin{center}
    \FigureFile(80mm,80mm){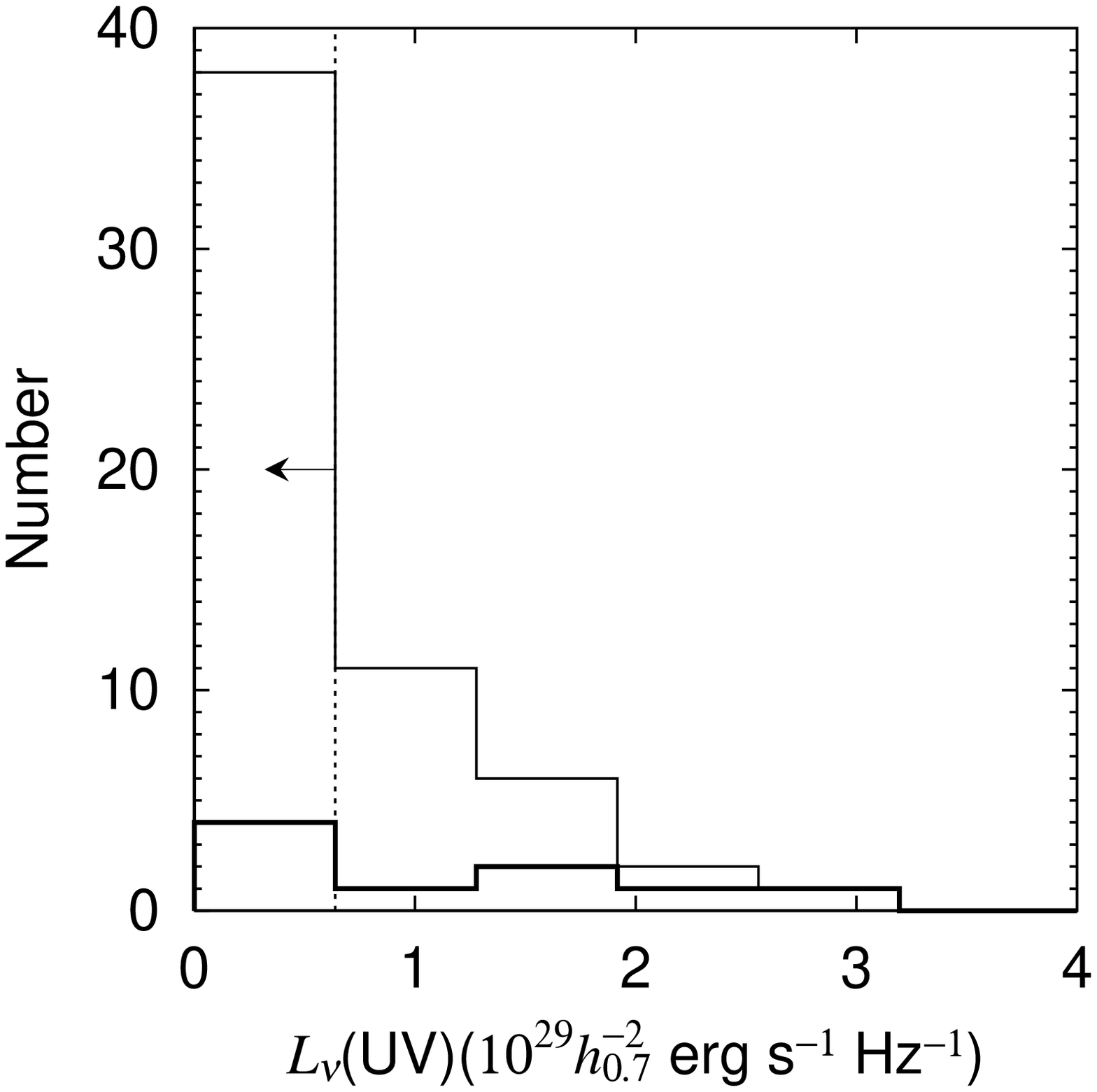}
  \end{center}
\caption{
Histograms of $L_\nu$(UV) at $\lambda$ = 1260 \AA ~ 
for the photometric sample (the 58 LAE
candidates) and for the nine LAEs. The latter is shown by the thick line.
The bins correspond to 1$\sigma$
($= 0.64  \times 10^{29} ~ h_{0.7}^{-2}$ erg s$^{-1}$ \AA$^{-1}$),
2$\sigma$, and 3$\sigma$ of $L_\nu$(UV);
see table 7 in more details.
}\label{fig:fig14}
\end{figure}

\clearpage

\begin{figure}
  \begin{center}
    \FigureFile(80mm,80mm){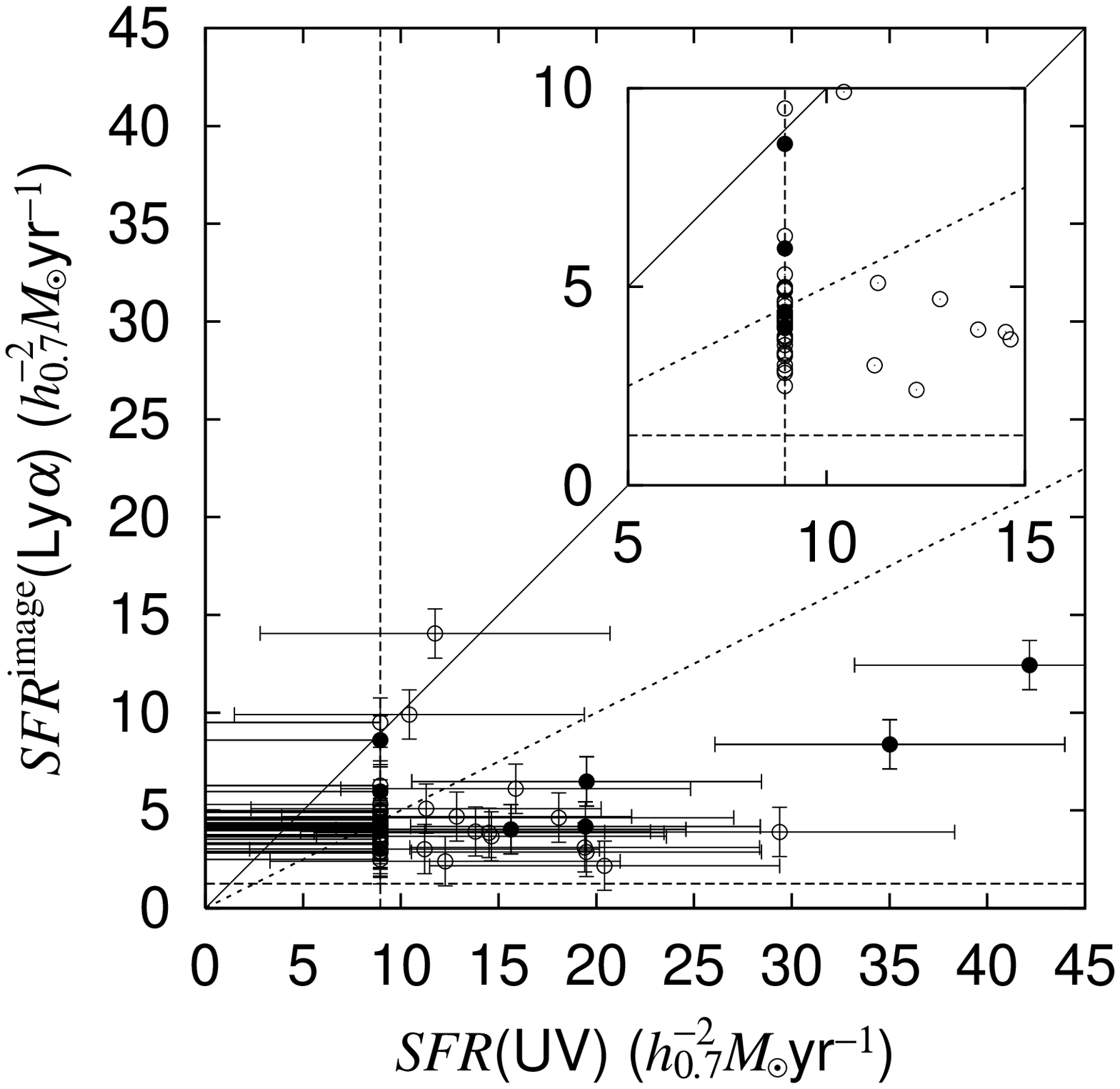}
  \end{center}
\caption{
Comparisons between $SFR^{\rm image}$(Ly$\alpha$) and $SFR$(UV)
for the 58 LAE candidates. The nine LAEs are shown by filled circles.
The inset shows a close up to see crowded data points for 
$SFR^{\rm image}$(Ly$\alpha$) $h_{0.7}^{-2} ~ M_\odot$ yr$^{-1}$ and for
$SFR$(UV) = 5$ -- $15 $h_{0.7}^{-2} ~ M_\odot$ yr$^{-1}$.
}\label{fig:fig15}
\end{figure}

\clearpage

\begin{figure}
  \begin{center}
    \FigureFile(80mm,80mm){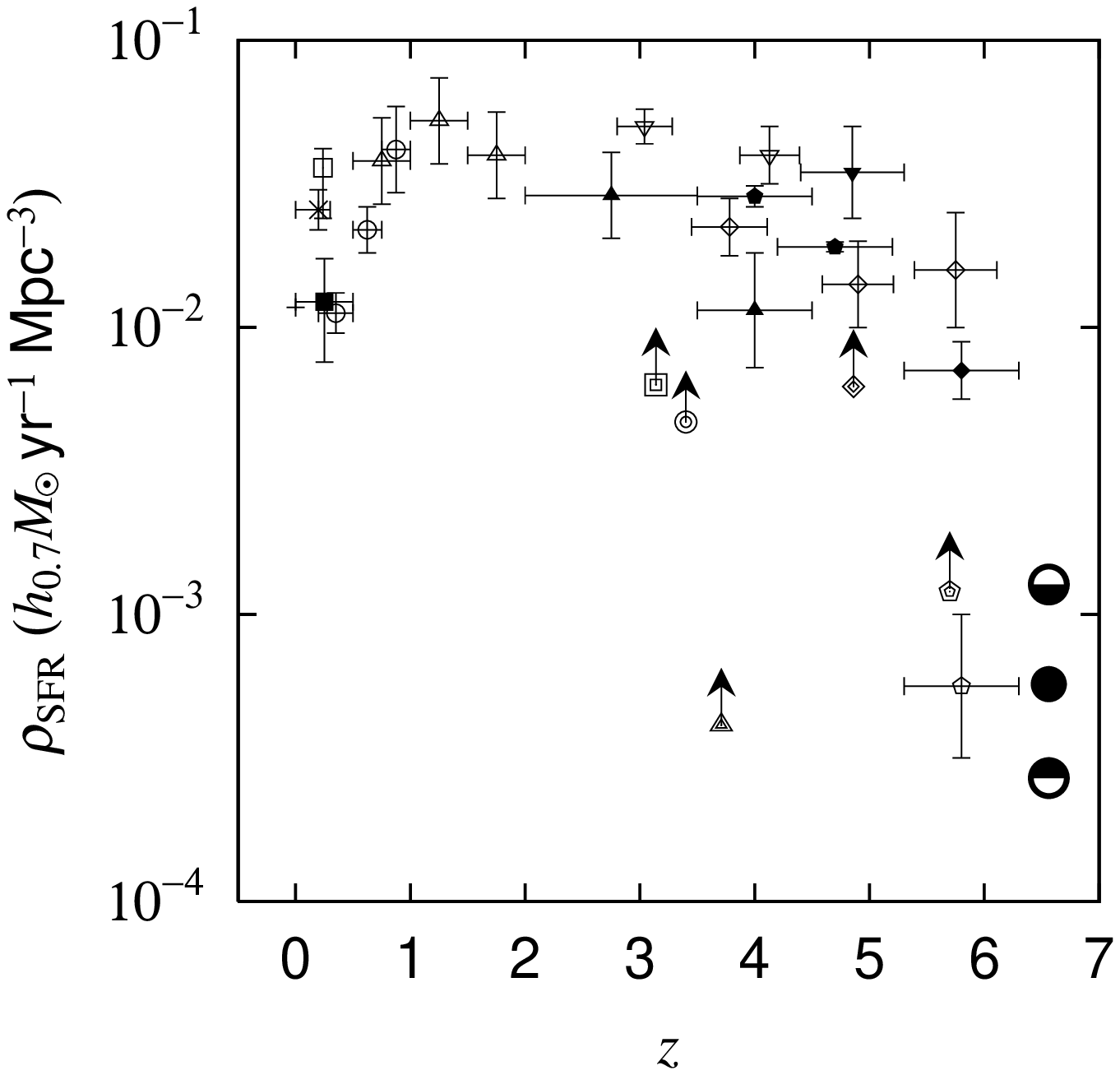}
  \end{center}
\caption{
Star-formation rate density shown as a 
function of the redshift. Our data points are shown
with big symbols; $\rho_{\rm SFR} = 5.5 \times 10^{-4}$ $h_{0.7}$ 
$M_\odot$ yr$^{-1}$  Mpc$^{-3}$ is
shown by the filled circle.  $\rho_{\rm SFR}^{\rm upper}$
is shown by an upper half-filled circle while $\rho_{\rm SFR}^{\rm lower}$
by an upper half-filled one.
The other data sources are
Gallego et al.    (1995 - plus),
Tresse and Maddox  (1998 - star),
Fujita et al.    (2003b - open square),
Treyer et al.     (1998 - filled square),
Lilly et al.      (1996 - open circles),
Connolly et al.   (1997 - open triangles),
Madau et al.      (1996 - filled triangles),
Steidel et al.    (1999 - open inverse triangles),
Iwata et al.      (2003 - filled inverse triangle),
Giavalisco et al. (2004b - open diamonds),
Bouwens et al.   (2004 - filled diamond),
Stanway et al.    (2004b - open hexagon), and
Ouchi et al.      (2004 - filled hexagons).
For reference, we also show
the results of previous Ly$\alpha$ searches at $z \sim 3$ -- 6 by
Kudritzki et al.  (2000 - double square),
Cowie and Hu       (1998 - double circle),
Fujita et al.     (2003a - double triangle),
Ouchi et al.      (2003 - double diamond),
and Ajiki et al.  (2003 - double hexagon).
}\label{fig:fig16}
\end{figure}


\begin{thebibliography}{}
\bibitem[]{}Ajiki, M., et al. 2002, ApJ, 576, L25
\bibitem[]{}Ajiki, M., et al. 2003, AJ, 126, 2091
\bibitem[]{}Bertin, E., \& Arnouts, S. 1996, A\&AS, 117, 393
\bibitem[]{}Bouwens, R. J., et al. 2004, ApJ, 606, L25
\bibitem[]{}Brocklehurst, M. 1971, MNRAS, 153, 471
\bibitem[]{}Bruzual, G., \& Charlot, S. 2003, MNRAS, 344, 1000
\bibitem[]{}Connolly, A. J., Szalay, A. S., Dickinson, M., Subbarao, M. U.,
            \& Brunner, R. J. 1997, ApJ, 486, L11
\bibitem[]{}Cowie, L. L., \& Hu, E. M. 1998, AJ, 115, 1319
\bibitem[]{}Dawson, S., Stern, D., Bunker, A. J., Spinrad, H., \&
            Dey, A. 2001, AJ, 122, 598
\bibitem[]{}Dickinson, M., et al. 2004, ApJ, 600, L99
\bibitem[]{}Ellis, R., Santos, M. R., Kneib, J. -P., \& Kuijken,
            K. 2001, ApJ, 560, L119
\bibitem[]{}Fioc, M., \& Rocca-Volmerange, B. 1997, A\&A, 326, 950
\bibitem[]{}Fujita, S. S., et al. 2003a, AJ, 125, 13
\bibitem[]{}Fujita, S. S., et al. 2003b, ApJ, 586, L115
\bibitem[]{}Gallego, J., Zamorano, J., Arag\'on-Salamanca, A., \& Rego, M.
            1995, ApJ, 455, L1; Errata 1996, 459, L43
\bibitem[]{}Giavalisco, M., et al. 2004a, ApJ, 600, L93
\bibitem[]{}Giavalisco, M., et al. 2004b, ApJ, 600, L103
\bibitem[]{}Haiman, Z. 2002, ApJ, 576, L1
\bibitem[]{}Hu, E. M., Cowie, L. L., Capak, P., McMahon, R. G., Hayashino, T.,
            \& Komiyama, Y.  AJ, 127, 563
\bibitem[]{}Hu, E. M., Cowie, L. L., McMahon, R. G., Capak, R., Iwamuro, F.,
            Kneib, J. -P., Maihara, T., \& Motohara, K. 2002, ApJ, 568, L75;
            Erratum, ApJ, 576, L99
\bibitem[]{}Iwata, I., Ohta, K., Tamura, N., Ando, M., Wada, S., Watanabe, C.,
            Akiyama, M., \& Aoki, K. 2003, PASJ, 55, 415
\bibitem[]{}Iye, M., et al. 2004, PASJ, 56, 381
\bibitem[]{}Kaifu, N., et al. 2000, PASJ, 52, 1
\bibitem[]{}Kashikawa, N., et al. 2002, PASJ, 54, 819
\bibitem[]{}Kashikawa, N., et al. 2003, AJ, 125, 53
\bibitem[]{}Kashikawa, N., et al. 2004a, PASJ, 56, 1011
\bibitem[]{}Kennicutt, R. C., Jr. 1998, ARA\&A, 36, 189
\bibitem[]{}Kneib, J. -P., Ellis, R. S.,  Santos, M. R., \& Richard, J.
            2004, ApJ, 607, 697
\bibitem[]{}Kodaira, K., et al. 2003, PASJ, 55, L17
\bibitem[]{}Kudritzki, R.-P., et al. 2000, ApJ, 536, 19
\bibitem[]{}Kurk, J. D., Cimatti, A., di Sereo Alighieri, S., Vernet, J.,
            Daddi, E., Ferrara, A., \& Ciardi, B. 2004, A\&A, 422, L13
\bibitem[]{}Lilly, S. J., Le Fevre, O., Hammer, F., \& Crampton, D. 1996,
            ApJ, 460, L1
\bibitem[]{}Madau, P., Ferguson, H. C., Dickinson, M. E., Giavalisco, M.,
            Steidel, C. C., \& Fruchter, A. 1996, MNRAS, 283, 1388
\bibitem[]{}Madau, P., Pozzetti, L., \& Dickinson, M. 1998, ApJ, 498, 106
\bibitem[]{}Maier, C., et al. 2003, A\&A, 402, 79
\bibitem[]{}Maihara, T., et al. 2001, PASJ, 53, 25
\bibitem[]{}Miyazaki, S., et al. 2002, PASJ, 54, 833
\bibitem[]{}Nagao, T., et al. 2004, ApJ, 613, L9
\bibitem[]{}Ostriker, J. P., \& Gnedin, N. Y. 1996, ApJ, 472, L63
\bibitem[]{}Ouchi, M., et al. 2001, BAAS, 33, 1313
\bibitem[]{}Ouchi, M., et al. 2003, ApJ, 582, 60 
\bibitem[]{}Ouchi, M., et al. 2004a, ApJ, 611, 660
\bibitem[]{}Ouchi, M., et al. 2004b, ApJ, 611, 685
\bibitem[]{}Pell\'o, R., Schaerer, D., Richard, J., Le Borgne, J. -F.,
            \& Kneib, J. -P. 2004a, A\&A, 416, L35
\bibitem[]{}Pell\'o, R.,  Richard, J., Le Borgne, J. -F., \& Schaerer, D.
            2004b, astro-ph/0407194
\bibitem[]{}Reddy, N. A., \& Steidel, C. C. 2004, ApJ, 603, L13
\bibitem[]{}Rhoads, J. E., et al. 2003, AJ, 125, 1006
\bibitem[]{}Rhoads, J. E., et al. 2004, ApJ, 611, 59
\bibitem[]{}Rhoads, J. E., \& Malhotra, S. 2001, ApJ, 563, L5
\bibitem[]{}Santos, M. R., Ellis, R. S., Kneib, J. -P., Richard, J., \&
            Kuijken, K. 2004, ApJ, 606, 683
\bibitem[]{}Shimasaku, K., et al. 2003, ApJ, 586, L111
\bibitem[]{}Shimasaku, K., et al. 2004, ApJ, 605, L93
\bibitem[]{}Shioya, Y., et al. 2002, PASJ, 54, 975
\bibitem[]{}Spinrad, H. 2003 Astrophysics Update, in press (astro-ph/0308411)
\bibitem[]{}Stanway, E. R., et al. 2004a, ApJ, 604, L13
\bibitem[]{}Stanway, E. R., McMahon, R. G., \& Bunker, A. 
            2004b, MNRAS, submitted (astro-ph/0403585)
\bibitem[]{}Steidel, C. C., Adelberger, K. L., Giavalisco, M., Dickinson, M., 
            Pettini, M. 1999, ApJ, 519, 1
\bibitem[]{}Steidel, C. C., Giavalisco, M., Pettini, M.,  Dickinson, M., \&
            Adelberger, K. L. 1996, ApJ, 462, L17
\bibitem[]{}Stern, D., Bunker, A., Spinrad, H., \& Dey, A. 2000,
            ApJ, 537, 73
\bibitem[]{}Stern, D., et al. 2004, ApJ, submitted (astro-ph/0407409)
\bibitem[]{}Taniguchi, Y. 2003, in the Proceedings of ^^ ^^ Multiwavelength Mapping
            of Galaxy Formation and Evolution", in press (astro-ph/0312228)
\bibitem[]{}Taniguchi, Y., et al. 2003a, ApJ, 585, L97
\bibitem[]{}Taniguchi, Y., Shioya, Y., Ajiki, M., Fujita, S. S., Nagao, T., \&
            Murayama, T.
            2003b, JKAS, 36, 123 (astro-ph/0306409); Erratum, JKAS, 36, 283
\bibitem[]{}Totani, T., Yoshii, Y., Iwamuro, F., Maihara, T., \& Motohara, K.
            2001a, ApJ, 550, L137
\bibitem[]{}Totani, T., Yoshii, Y., Iwamuro, F., Maihara, T., \& Motohara, K.
            2001b, ApJ, 558, L87
\bibitem[]{}Totani, T., Yoshii, Y., Maihara, T., Iwamuro, F., \& Motohara, K.
            2001c, ApJ, 559, 592
\bibitem[]{}Tresse, L., \& Maddox, S. J. 1998, ApJ, 495, 691
\bibitem[]{}Treyer, M. A., Ellis, R. S., Milliard, B., Donas, J., \& Bridges,
            T. J. 1998, MNRAS, 300, 303
\bibitem[]{}Weatherley, S. J., Warren, S. J., \& Babbedge, T. S. R. 2004,
            A\&A, 428, L29
\bibitem[]{}Wyithe, J. S. B., \& Loeb, A. 2002, ApJ, 577, 57
\bibitem[]{}Yagi M.,  Kashikawa, N., Sekiguchi, M., Doi, M., Yasuda, N.,
              Shimasaku, K., \& Okamura, S. 2002, AJ, 123, 66
\bibitem[]{}Yamada, S. F., et al. 2003, PASJ, 55, 733
\end{thebibliography}
\end{document}